\documentclass{pj}
\usepackage{graphicx}
\usepackage{cite}

\begin{document}
\setcounter{page}{1}
\pjheader{Vol.\ x, y--z, 2014}

\title[Running head]
{Optimized Superconducting Nanowire Single Photon Detectors\\*to Maximize Absorptance}
 \footnote{\it Received date}  \footnote{\hskip-0.12in*\, Corresponding
author:~M\'{a}ria~Csete (mcsete@physx.u-szeged.hu).}
\footnote{\hskip-0.12in\textsuperscript{1} Department of Optics and Quantum Electronics, University of Szeged, D\'{o}m t\'{e}r 9, Szeged, H-6720, Hungary.\\ 
\textsuperscript{2} Institute of Informatics, University of Szeged, \'{A}rp\'{a}d t\'{e}r 2, Szeged, H-6720, Hungary.}

\author{M\'{a}ria~Csete\textsuperscript{*1}, G\'{a}bor~Szekeres\textsuperscript{1}, Andr\'{a}s~Szenes\textsuperscript{1}, Bal\'{a}zs~B\'{a}nhelyi\textsuperscript{2},\\*Tibor~Csendes\textsuperscript{2} and G\'{a}bor~Szab\'{o}\textsuperscript{1}}

\runningauthor{Csete et al.}

\tocauthor{FistName1~LastName1 and FistName1~LastName1}

\begin{abstract}
Dispersion characteristics of four types of superconducting nanowire single photon detectors, nano-cavity-array- (NCA-), nano-cavity-deflector-array- (NCDA-), nano-cavity-double-deflector-array- (NCDDA-) and nano-cavity-trench-array- (NCTA-) integrated (I-A-SNSPDs) devices was optimized in three periodicity intervals commensurate with half-, three-quarter- and one SPP wavelength.~The optimal configurations capable of maximizing NbN absorptance correspond to periodicity dependent tilting in S-orientation (90$^{\circ}$ azimuthal orientation).~In NCAI-A-SNSPDs absorptance maxima are reached at the plasmonic Brewster angle (PBA) due to light tunneling. The absorptance maximum is attained in a wide plasmonic-pass-band in NCDAI$_{1/2*\lambda}$-A, inside a flat-plasmonic-pass-band in NCDAI$_{3/4*\lambda}$-A and inside a narrow plasmonic-band in NCDAI$_{\lambda}$-A. In NCDDAI$_{1/2*\lambda}$-A bands of strongly-coupled cavity and plasmonic modes cross, in NCDDAI$_{3/4*\lambda}$-A an inverted-plasmonic-band-gap develops, while in NCDDAI$_{\lambda}$-A a narrow plasmonic-pass-band appears inside an inverted-minigap. The absorptance maximum is achieved in NCTAI$_{1/2*\lambda}$-A inside a plasmonic-pass-band, in NCTAI$_{3/4*\lambda}$-A at inverted-plasmonic-band-gap center, while in NCTAI$_{\lambda}$-A inside an inverted-minigap. The highest 95.05\% absorptance is attained at perpendicular incidence onto NCTAI$_{\lambda}$-A. Quarter-wavelength type cavity modes contribute to the near-field enhancement around NbN segments except in NCDAI$_{\lambda}$-A and NCDDAI$_{3/4*\lambda}$-A. The polarization contrast is moderate in NCAI-A-SNSPDs ($\sim10^2$), NCDAI- and NCDDAI-A-SNSPDs make possible to attain considerably large polarization contrast ($\sim10^2-10^3$ and $\sim10^3-10^4$), while NCTAI-A-SNSPDs exhibit a weak polarization selectivity ($\sim10-10^2$).
\end{abstract}


\setlength {\abovedisplayskip} {6pt plus 3.0pt minus 4.0pt}
\setlength {\belowdisplayskip} {6pt plus 3.0pt minus 4.0pt}

\

\section{Introduction}
\label{introduction}

Single-photon generation and detection are key steps of quantum information processing (QIP) \cite{article01, article02, article03, article04, article05, article06, article07, article08}. The main characteristic parameters of single-photon detectors include detection efficiency, dark-count rate, reset time and timing jitter. Superconducting materials are widely used in single-photon detectors, e.g. in superconducting nanowire single-photon detectors (SNSPD) \cite{article01, article02, article03, article04, article05, article06, article07, article08, article09, article10, article11, article12, article13, article14, article15, article16, article17, article18, article19, article20, article21, article22, article23, article24, article25, article26, article27, article28, article29, article30, article31, article32, article33}. The system detection efficiency of SNSPDs is determined as $SDE= \eta_{external}*\eta_{absorption}*\eta_{registering}$, where $\eta_{external}$ qualifies the efficiency of coupling from free space, $\eta_{absorption}$ is the absorption efficiency determined by geometrical and optical properties of the detector, and $\eta_{registering}$ indicates the probability that the incident photon generates a voltage signal. There are several efforts described in the literature to improve SNSPDs performance by tailoring the superconducting pattern properties. The $\eta_{registering}$ efficiency was enhanced via ultra-narrow superconducting NbN wires \cite{article22}, and by applying novel superconducting materials \cite{article19, article28, article29}. It was shown that meandered superconducting patterns with rounded corners make it possible to enhance the critical current and to reduce dark current \cite{article27}.

According to Ginzberg-Landau theory shorter nanowires possess smaller kinetic inductance causing shorter reset time \cite{article09}, therefore various nanophotonical methodologies were developed to enhance the absorption efficiency via short meandered superconducting patterns. Cavities have been primarily applied to enhance the $\eta_{external}*\eta_{absorption}$ effective absorption cross-section \cite{article10, article13, article18, article19, article23, article25}. Distributed Bragg reflectors and multi-layers acting as optical cavities were also implemented \cite{article21, article29}. Plasmonic structures capable of improving light-in-coupling via localized and propagating modes were recently integrated into SNSPDs, and large enhanced effective absorption cross-section qualified by $\eta_{plasmonic}*\eta_{absorption}$ quantity was achieved \cite{article24, article26, article31, article32, article33}. Moreover, it was shown that the plasmonic nature of superconducting stripes can be also used to realize efficient single-photon detection \cite{article30}. 

Stripes embedded into dielectric media and into optical cavities inherently prefer \textbf{E}-field oscillation parallel to their long axes \cite{article13, article23, article25}. To quench the corresponding polarization sensitivity of detectors, spiral geometry was proposed \cite{article14}, and two mutually perpendicular patterns were stacked vertically into a multilayer cavity \cite{article28}. The polarization sensitivity of superconducting patterns located in close proximity of antennas and embedded into complex plasmonic structures differs fundamentally from that of bare stripes \cite{article24, article26, article31, article32, article33}. Superconducting patterns integrated with 1D periodic plasmonic structures of different types exhibit enhanced absorptance at azimuthal orientations corresponding the \textbf{E}-field oscillation perpendicular to metal segments \cite{article26, article31, article33}. Based on our previous results the p-polarized absorptance can be significantly improved with respect to the s-polarized one, therefore plasmonic structure integrated devices are particularly promising in QIP applications, where polarization selectivity is very important, e.g. in Bennett and Bassard QKD protocol (BB84) \cite{article26, article31, article33, article34, article35}. Detector designs ensuring QI specific read-out are crucial to avoid different attack schemes and in development of optical quantum computers \cite{article07, article08, article36, article37}.

All previous examples in the literature about SNSPD improvement are based on preconceptions regarding the structure parameters that are capable of enhancing detection efficiency. However, complete numerical optimization of integrated device structures have not been realized up until now. The main purpose of our present work was to determine those configurations of four different types of plasmonic structure integrated SNSPDs that are capable of maximizing the absorptance at 1550 nm wavelength, as well as to analyze the polarization contrast that can be achieved via these A-SNSPD type devices. In addition to this we have analyzed tendencies in configurations of four corresponding C-SNSPD type devices, which make possible to maximize the polarization contrast by gradually decreasing the criterion regarding the absorptance that have to be met parallel. These results are presented in Appendix (Figure~\ref{figA1}-\ref{figA4}).

\section{Methods}
\label{method}

Four different types of plasmonic structure integrated SNSPDs were inspected theoretically, and for each design types three periodicity regions were considered, where special nanophotonical phenomena are at play (Figure~\ref{fig1}). Taking into account the $\lambda_{SPP}$ wavelength of the surface plasmon polaritons (SPPs) excitable at silica substrate and gold interface at 1550 nm, periodicity regions commensurate with $0.5*\lambda_{SPP} / 0.75*\lambda_{SPP} / \lambda_{SPP}$ were considered, where Bragg scattering/extraordinary transmission/Rayleigh phenomenon are expected.

In nano-cavity-array integrated NCAI-A-SNSPDs each NbN stripe is surrounded by vertical gold segments composing an MIM nano-cavity-grating \cite{article26, article31, article33}. In nano-cavity-deflector-array integrated NCDAI-A-SNSPDs additional gold segments nominated as deflectors are inserted into the silica substrate at the anterior sides of each NbN loaded nano-cavity \cite{article31, article33}. In nano-cavity-double-deflector-array integrated NCDDAI-A-SNSPDs, both at their anterior and exterior sides, each nano-cavity is neighbored by gold deflectors, which can have different length and width \cite{article33}. In contrast, in nano-cavity-trench-array integrated NCTAI-A-SNSPDs trenches are embedded into the in-plane interleaved vertical gold segments, which can have different width \cite{article33, article38}. Insertion of trenches, which are also capable of acting as efficient plasmonic mirrors, makes it possible to reduce the total gold amount, which causes a competitive absorption.

Theoretical studies were performed to determine the optimal configurations for each type of SNSPD designs, using the special finite element method we have previously developed based on the Radio Frequency module of COMSOL Multiphysics software package (COMSOL AB) \cite{article23, article25, article26, article31, article33}. The index of refraction of dielectric materials (silica, NbNO$_{x}$ and HSQ) was specified via Cauchy formulae, while dielectric constants for both absorbing materials (NbN, Au) were loaded from tabulated datasets. 

We have applied an in-house developed efficient optimization technique, nominated as GLOBAL, for the solution of all problems emerging during SNSPD optimization, which was implemented using LiveLink for MATLAB in COMSOL. GLOBAL is a stochastic technique that is a sophisticated composition of sampling, clustering, and local search \cite{article39}. GLOBAL was also used successfully for the solution of very complex optimization problems, such as the mathematical proof of the chaotic behavior of the forced damped pendulum \cite{article40} and for proving a long standing conjecture of Wright on delay differential equations \cite{article41}.

Three-dimensional models were used to determine the optimal structure and the corresponding optimal illumination direction, which are capable of resulting in maximal absorptance, i.e. the optimal configuration of A-SNSPD type devices. All geometrical parameters and the $\varphi$ polar angle were varied at fixed $\gamma$ azimuthal angle (nominated as S-orientation), which results in the largest achievable absorptance in case of different plasmonic structure integrated SNSPDs, based on our previous studies \cite{article26, article31, article33}.

The dispersion characteristics in NbN absorptance was determined at the optimal $\gamma$ azimuthal orientation of four integrated A-SNSPD device types for three periodicity intervals as well, and is presented in Figure~\ref{fig2}. The angle-dependent absorptance of the optimized SNSPDs was analyzed first for 1550 nm p-polarized light illumination, than the wavelength dependency at the absorptance maxima was also inspected to uncover the underlying nanophotonical phenomena (Figure~\ref{fig2} and Figure~\ref{fig4}). 

The time-averaged \textbf{E}-field distribution was studied along with the power-flow at the maxima in polar angle on the NbN absorptance, at plane cross-sections taken perpendicular to single unit cells of the integrated SNSPDs (Figure~\ref{fig3}). The \textbf{E}-field time-evolution was inspected as well to characterize all localized and propagating modes supported by the integrated periodic patterns (see Multimedia files 1-12). We have inspected the ratio of absorptances achievable via p- and s-polarized light illumination in S-orientation. The polar-angle and wavelength-dependent polarization contrast was also analyzed for integrated A-SNSPD type devices (Figure~\ref{fig4}).

\section{Results}
\label{results}

The optimization results indicated that the optimal configurations are device design specific, and the detailed inspection of the optimized A-SNSPD type devices' dispersion characteristics and near-field distribution revealed that the underlying nanophotonics fundamentally depends on the integrated structure type and on the periodicity interval.

\subsection{Parameters and Dispersion Characteristics of Optimized A-SNSPD Configurations}
\label{param and disp}

\begin{figure}[h]
\centerline{\includegraphics[width=1\columnwidth,draft=false]{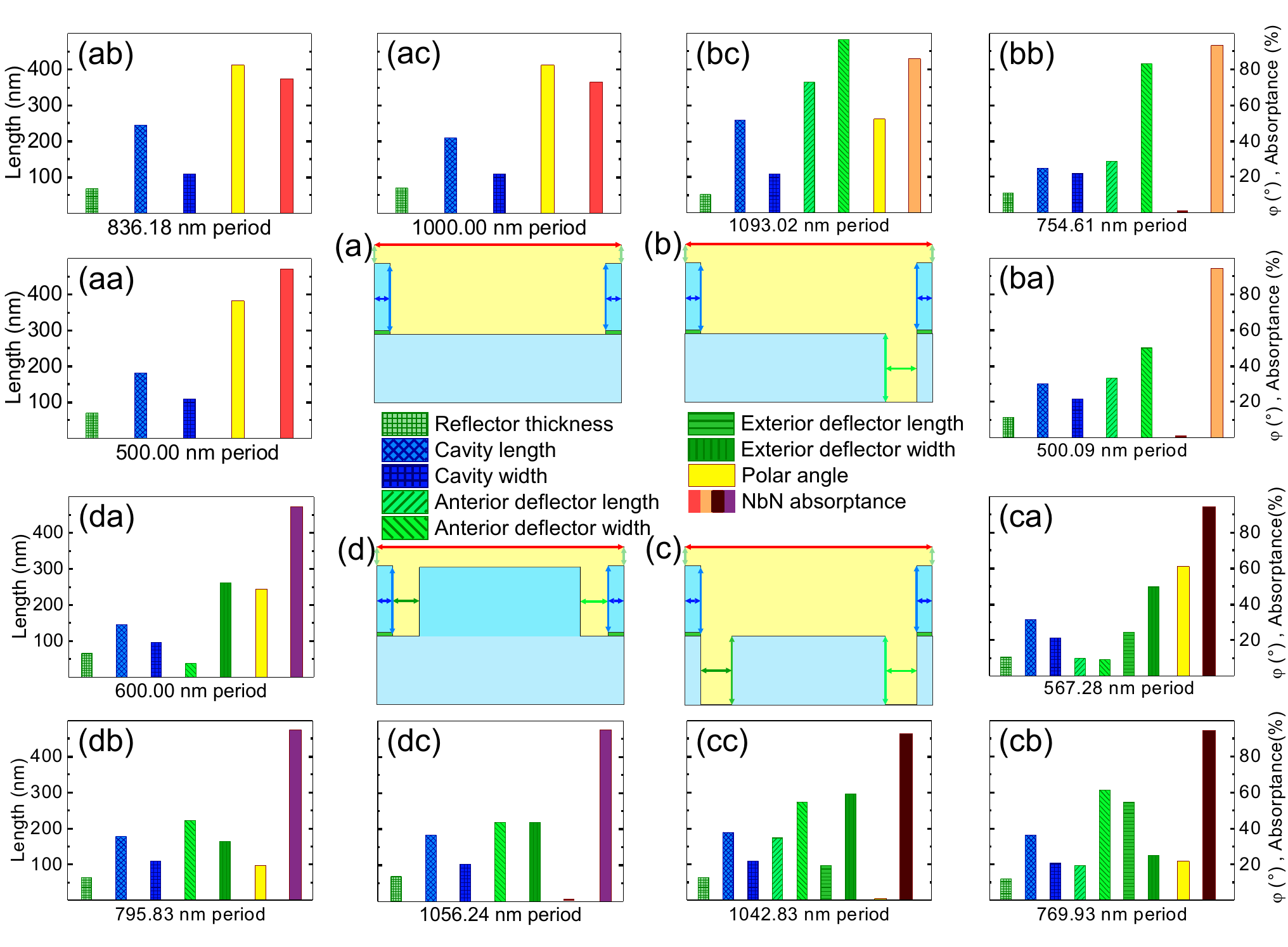}}
\caption{Schematics of geometry optimization method of (a) NCAI-, (b) NCDAI-, (c) NCDDAI- and (d) NCTAI-SNSPD. Histograms showing the optimized geometrical parameters, absorptance and optimal polar angle in (aa) NCAI$_{1\backslash2*\lambda}$-A, (ab) NCAI$_{3\backslash4*\lambda}$-A, (ac) NCAI$_{\lambda}$-A, (ba) NCDAI$_{1\backslash2*\lambda}$-A, (bb) NCDAI$_{3\backslash4*\lambda}$-A, (bc) NCDAI$_{\lambda}$-A, (ca) NCDDAI$_{1\backslash2*\lambda}$-A, (cb) NCDDAI$_{3\backslash4*\lambda}$-A, (cc) NCDDAI$_{\lambda}$-A, (da) NCTAI$_{1\backslash2*\lambda}$-A, (db) NCTAI$_{3\backslash4*\lambda}$-A and (dc) NCTAI$_{\lambda}$-A.}
\label{fig1}
\end{figure}

The optimization of half-wavelength-scaled integrated NCAI$_{1/2*\lambda}$-A-SNSPD pattern resulted in 500 nm optimal pitch and 110 nm optimal MIM nano-cavity width, which corresponds to the lower and upper bound of the inspected periodicity and nano-cavity width interval, respectively (Figure~\ref{fig1}(aa)). This suggests that allowing smaller periodicities would result in even higher absorptance, however at the expense of kinetic inductance and reset time increase, therefore we have finalized the optimization procedure at this pitch. The nano-cavity length is 181.44 nm in the optimized NCAI$_{1/2*\lambda}$-A device (Figure~\ref{fig1}(aa)).

\begin{figure}[h]
\centerline{\includegraphics[width=1\columnwidth,draft=false]{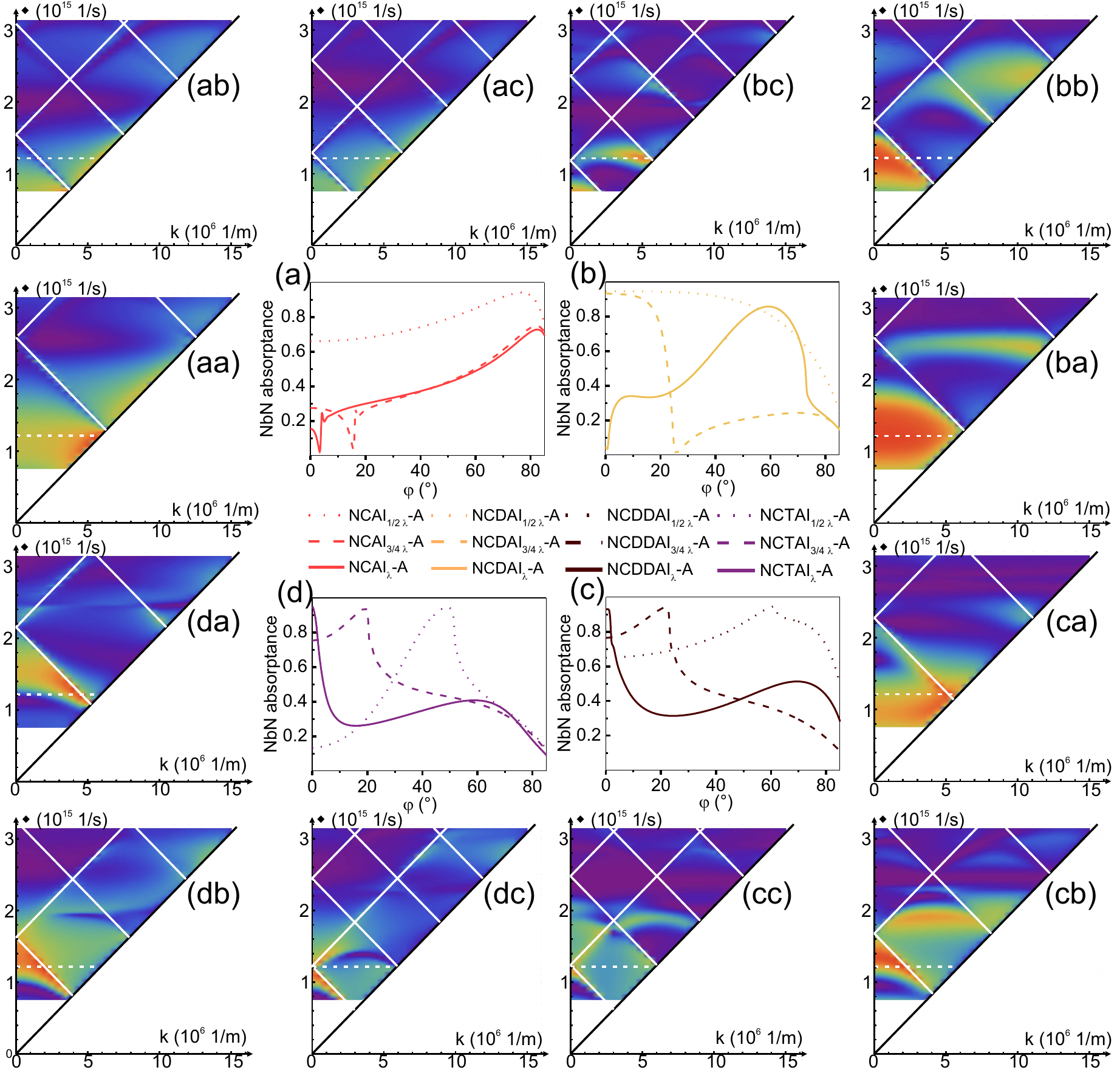}}
\caption{Polar angle dependent absorptances of (a) NCAI-, (b) NCDAI-, (c) NCDDAI- and (d) NCTAI-SNSPD in S-orientation. Dispersion diagrams of (aa) NCAI$_{1\backslash2*\lambda}$-A, (ab) NCAI$_{3\backslash4*\lambda}$-A, (ac) NCAI$_{\lambda}$-A, (ba) NCDAI$_{1\backslash2*\lambda}$-A, (bb) NCDAI$_{3\backslash4*\lambda}$-A, (bc) NCDAI$_{\lambda}$-A, (ca) NCDDAI$_{1\backslash2*\lambda}$-A, (cb) NCDDAI$_{3\backslash4*\lambda}$-A, (cc) NCDDAI$_{\lambda}$-A, (da) NCTAI$_{1\backslash2*\lambda}$-A, (db) NCTAI$_{3\backslash4*\lambda}$-A and (dc) NCTAI$_{\lambda}$-A at global absorptance maxima.}
\label{fig2}
\end{figure}

The dispersion diagram of the optimized NCAI$_{1/2*\lambda}$-A shows that the absorptance is enhanced throughout almost the entire polar angle interval in a wide spectral region around 1550 nm (Figure~\ref{fig2}(aa)). Moreover, the NbN absorptance is locally enhanced at tilting corresponding to the wavelength dependent plasmonic Brewster angle (PBA) close to the first Brillouin zone boundary \cite{article31, article33, article42, article43, article44}. The collective resonance on the half-wavelength-scaled nano-cavity-array results in slightly polar angle dependent absorptance (Figure~\ref{fig2}(a)), according to previous observations on short-pitch gratings \cite{article45, article46, article47, article48}. Namely, the absorptance is 65.97\% already at perpendicular incidence and increases monotonically. The 94.18\% global maximum is attained at 76.38$^{\circ}$ tilting, which corresponds to the PBA of the NCAI$_{1/2*\lambda}$-A.

The optimization of three-quarter-wavelength-scaled NCAI$_{3/4*\lambda}$-A-SNSPD integrated devices resulted in 836.18 nm optimal periodicity. The 109.97 nm nano-cavity width is almost equal to corresponding optimal parameter of NCAI$_{1/2*\lambda}$-A, while the 243.79 nm nano-cavity length is significantly larger (Figure~\ref{fig1}(ab)). The dispersion graph of the optimized NCAI$_{3/4*\lambda}$-A device shows that considerably smaller absorptance is attainable inside a significantly smaller polar angle and wavelength interval (Figure~\ref{fig2}(ab)). In addition to this, the NCAI$_{3/4*\lambda}$-A is capable of coupling the incident light into surface waves at small tilting, which causes a significant modulation on the dispersion curve \cite{article49, article50, article51, article52, article53}. The absorption maxima appear close to the second Brillouin zone boundary. At 1550 nm the light in-coupling results in a 3.75\% global absorptance minimum at 15.60$^{\circ}$ polar angle (Figure~\ref{fig2}(a)), which is followed by a local absorptance maximum - minimum pair resulting in 27.35\% and 21.82\% absorptance at 16.50$^{\circ}$ and 17.30$^{\circ}$ tilting, respectively. These features are in accordance with Wood-anomaly \cite{article31, article33, article53, article54, article55, article56}. By increasing the polar angle a monotonic NbN absorptance increase is observable through the PBA \cite{article31, article33, article42, article43, article44}. The light tunneling phenomena result in a global absorptance maximum of 74.96\% at 82.24$^{\circ}$ tilting.

The optimization of wavelength-scaled NCAI$_{\lambda}$-A-SNSPD resulted in maximal NbN absorptance in 1000 nm pitch pattern of 110 nm width MIM nano-cavities, which corresponds to the lower and upper bound of periodicity and cavity width intervals, respectively (Figure~\ref{fig1}(ac)). The highest absorptance is achieved via nano-cavities with intermediate 208.82 nm optimal length. The dispersion characteristics is similar to that of NCAI$_{3/4*\lambda}$-A, and exhibits a grating coupling related modulation \cite{ article49, article50, article51, article52, article53}. The absorptance modulation caused by coupling to surface modes appears at smaller polar angles according to the increased periodicity (Figure~\ref{fig2}(ac)) \cite{article31, article33, article53, article54, article55, article56}. The largest absorptance is achieved close to the second Brillouin zone boundary. At 1550 nm 2.15\% global minimum appears at 3.40$^{\circ}$ tilting, which is followed by a local maximum-minimum pair, manifesting itself in 24.51\% and 19.63\% NbN absorptance at 4.2$^{\circ}$ and 5.1$^{\circ}$ tilting. Then the NbN absorptance monotonously increases through the 72.82\% global absorptance maximum appearing at 82.46$^{\circ}$ tilting corresponding to the PBA (Figure~\ref{fig2}(a)) \cite{article31, article33, article42, article43, article44}. The attained absorptance is smaller than the absorptance observed in either of the NCAI$_{1/2*\lambda}$-A and NCAI$_{3/4*\lambda}$-A devices almost in the entire polar angle interval. 

The optimization of half-wavelength-scaled NCDAI$_{1/2*\lambda}$-A-SNSPD resulted in 500.09 nm optimal periodicity and 108.72 nm optimal nano-cavity width (Figure~\ref{fig1}(ba)). Their noticeable difference with respect to lower/upper optimization bounds proves that the parameters correspond to an efficiently optimized design. The 149.89 nm nano-cavity length in the optimized device is even smaller than in case of NCAI$_{1/2*\lambda}$-A. The optimized 166.46 nm length of deflectors is slightly larger, and in spite of their large 250.98 nm width, large NbN absorptance is attained. The dispersion diagram of the optimized NCDAI$_{1/2*\lambda}$-A shows that this integrated device supports collective resonance in a plasmonic pass band covering the widest spectral and polar angle interval (Figure~\ref{fig2}(ba)), and almost polar angle independent absorptance is achieved according to the sub-wavelength pitch \cite{article45, article46, article47, article48}. The absorptance maxima appear close to perpendicular incidence in a wide spectral region, than the absorptance decreases monotonously towards the first Brillouin zone boundary. The slightly polar angle dependent absorptance of NCDAI$_{1/2*\lambda}$-A shows a 94.68\% global maximum close to perpendicular incidence at 0.04$^{\circ}$ tilting. The absorptance decreases monotonically also at 1550 nm revealing the absence of PBA corresponding to NbN loaded nano-cavities. 

The optimization of three-quarter-wavelength-scaled integrated NCDAI$_{3/4*\lambda}$-A-SNSPD integrated device resulted in 754.62 nm periodicity close to the lower bound. The 109.61 nm nano-cavity width is slightly larger than in NCDAI$_{1/2*\lambda}$-A. The 124.49 nm and 144.55 nm nano-cavity and deflector lengths are slightly smaller but commensurate with those in NCDAI$_{1/2*\lambda}$-A. In contrast to intuitive expectations the 415.24 nm optimized deflector width is considerably larger than the width of deflectors in NCDAI$_{1/2*\lambda}$-A (Figure~\ref{fig1}(bb)). The dispersion graph of the optimized NCDAI$_{3/4*\lambda}$-A device shows that in presence of deflectors large absorption is achieved through significant fraction of the first Brillouin zone. The largest absorptance is attainable already at perpendicular incidence in wide spectral interval inside a flat plasmonic pass band, which is noticeably smaller than in NCDAI$_{1/2*\lambda}$-A (Figure~\ref{fig2}(bb)). The NCDAI$_{3/4*\lambda}$-A is capable of coupling the incident light into surface waves at regions of small transitional tilting with large efficiency. The resulted flat pass-band on the dispersion curve is followed by a well-defined cut-off, where the first order grating coupling occurs \cite{ article49, article50, article51, article52, article53}. The absorptance of NCDAI$_{3/4*\lambda}$-A is 93.34\% at perpendicular incidence (Figure~\ref{fig2}(b)), which is the global maximum of this device. The absorptance decreases by increasing the polar angle, at 25.50$^{\circ}$ polar angle reaches 0.76\%, i.e. it is almost zero according to the cut-off on the integrated structure. This orientation corresponds to the first Brillouin zone boundary, where the integrated pattern couples in -1 order into backward propagating Brewster-Zenneck modes in the composite diffracted field. These modes have a 1084.42 wavelength, which is larger than the wavelength of photonic modes \cite{ article49, article50, article51, article52, article53, article57, article58, article59, article60}. For larger polar angles the NbN absorptance increases slowly again. However, only a weak absorptance enhancement is observable in the region, where PBA phenomenon originating from array of NbN loaded cavities is expected.

The optimization of wavelength-scaled NCDAI$_{\lambda}$-A-SNSPD resulted in the largest absorptance in 1093.02 nm pitch integrated pattern consisting of 109.41 nm wide nano-cavities, which parameters are closer to the upper bound (Figure~\ref{fig1}(bc)). The 258.02 nm nano-cavity length is considerably larger than in corresponding NCDAI$_{1/2*\lambda}$-A and NCDAI$_{3/4*\lambda}$-A devices. The deflectors having 364.02 nm length and 481.46 nm width are the longest and widest among all optimized SNSPD devices. The dispersion characteristics significantly differs from that of the NCDAI$_{1/2*\lambda}$-A and NCAI$_{3/4*\lambda}$-A devices, and indicates a weak and narrow plasmonic band (Figure~\ref{fig2}(bc)) \cite{ article49, article50, article51, article52, article53}. The largest NbN absorptance is achieved close to the second Brillouin zone boundary due to grating coupling in -2 order into backward propagating modes. The absorptance characteristics of NCDAI$_{\lambda}$-A at 1550 nm significantly differs from that of two other NCDAI-A-SNSPD devices. Namely, the NbN absorptance is 4.5\% at perpendicular incidence, than increases rapidly until 5.00$^{\circ}$ polar angle, and after an inflection point further significant increase is observable. The 85.77\% global absorption maximum appears at 59.25$^{\circ}$ tilting, which polar angle is significantly smaller than the PBA corresponding to the array of NbN loaded nano-cavities. The 1093.02 nm pitch grating couples in -2 order at this tilting into surface modes having a 975.45 nm wavelength, which is significantly smaller than the photonic wavelength, and indicates cavity and propagating modes' interaction \cite{article26, article31, article33, article49, article50}. The attained absorptance is smaller than in NCDAI$_{3/4*\lambda}$-A through $\sim25.00^{\circ}$ polar angle, while it is overridden by the absorptance achieved in NCDAI$_{1/2*\lambda}$-A in the entire polar angle interval. 

The optimization of half-wavelength-scaled NCDDAI$_{1/2*\lambda}$-A-SNSPD resulted in 567.82 nm optimal periodicity and 106.77 nm nano-cavity width, which are close to the upper bounds (Figure~\ref{fig1}(ca)). The 157.69 nm nano-cavity length is intermediate compared to those of NCAI$_{1/2*\lambda}$-A and NCDAI$_{1/2*\lambda}$-A. The 50 nm/123.26 nm lengths and 46.36 nm/248.79 nm widths reveal that gold deflectors at the anterior/exterior side of nano-cavities play negligible/dominant role (see Appendix Figure~\ref{figA3}(ca)). According to the dispersion diagram of NCDDAI$_{1/2*\lambda}$-A the desired 1550 nm is at the upper edge of a strongly-coupled region of a collectively resonant nano-cavity mode and a backward propagating surface mode originating from -1 order grating coupling (Figure~\ref{fig2}(ca)) \cite{article50, article62}. The strong-coupling phenomenon results in local enhancement inside a characteristic cross-shaped area proving mode-hybridization. The absorptance is slightly polar angle dependent in NCDDAI$_{1/2*\lambda}$-A according to collective resonances on the sub-wavelength nano-cavity-array \cite{article45, article46, article47, article48}. The NbN absorptance is 65.00\% at perpendicular incidence and monotonically increases with tilting until a modulation, which results in a 94.60\% global absorptance maximum at 60.89$^{\circ}$. The 1060.19 nm wavelength of the backward propagating modes coupled in -1 order indicates their plasmonic nature \cite{article31, article33}. The absorptance decreases monotonously towards the first Brillouin zone boundary indicating that light tunneling phenomenon at the PBA originating from array of NbN loaded nano-cavities is depressed also in presence of double deflectors.

The optimization of three-quarter-wavelength scaled NCDDAI$_{3/4*\lambda}$-A-SNSPD resulted in 769.93 nm periodicity and 102.24 nm nano-cavity width, which are close to the lower bound (Figure~\ref{fig1}(cb)). In NCDDAI$_{3/4*\lambda}$-A the smaller cavity width is accompanied by larger, 182.15 nm length compared to NCDDAI$_{1/2*\lambda}$-A. The 97.04 nm/273.12 nm ratio of deflector lengths at anterior/exterior sides is slightly decreased with respect to that observed in NCDDAI$_{1/2*\lambda}$-A, while the ratio of their 307.43 nm/124.13 nm widths is reversed, which reveals that the two gold deflectors play more compensated role (see Appendix Figure~\ref{figA3}(cb)). The dispersion graph of the optimized NCDDAI$_{3/4*\lambda}$-A device indicates a plasmonic pass band, where large absorptance is attainable at transitional tilting in an intermediate spectral interval (Figure~\ref{fig2}(cb)) \cite{ article49, article50, article51, article52, article53}. The NCDDAI$_{3/4*\lambda}$-A is capable of coupling the incident light in -1 order into backward propagating surface modes at tilting similar to the polar angle corresponding to cut-off in NCDAI$_{3/4*\lambda}$-A. This indicates that double deflectors evolve phase correction effect on the coupled propagating modes \cite{article33}. The largest absorption is achieved close to the top of the first Brillouin zone inside an inverted plasmonic band-gap. NCDDAI$_{3/4*\lambda}$-A shows an enhanced 76.39\% absorptance already at perpendicular incidence. By increasing tilting the absorptance increases until 21.85$^{\circ}$ polar angle, where 94.34\% global maximum is achieved due to coupling in -1 order into backward propagating surface modes having a wavelength of 1052.13 nm. This wavelength indicates that plasmonic modes are coupled at the inverted PBG center. By increasing tilting further the absorptance does not indicate enhancement at PBA corresponding to the array of NbN loaded nano-cavities. 

The optimization of wavelength-scaled NCDDAI$_{\lambda}$-A-SNSPD resulted in maximal absorptance in 1042.83 nm pitch integrated pattern consisting of 109.15 nm wide nano-cavities (Figure~\ref{fig1}(cc)). Although, the optimal 190.06 nm nano-cavity length is again larger than in NCDDAI$_{1/2*\lambda}$-A and NCDDAI$_{3/4*\lambda}$-A, it is smaller than the optimal nano-cavity length in counterpart NCDAI$_{\lambda}$-A. The deflectors at the anterior/exterior side of the cavities have larger/smaller length, compared to NCDDAI$_{1/2*\lambda}$-A and NCDDAI$_{3/4*\lambda}$-A (see Appendix Figure~\ref{figA3}(cc)). The large 173.89 nm/97.32 nm lengths and 273.48 nm/295.77 nm widths at the anterior/exterior sides result in large competitive gold volume fraction. The dispersion characteristics indicates that not only the achieved absorptance, but also the extension of corresponding frequency-polar angle region is the smallest in NCDDAI$_{\lambda}$-A (Figure~\ref{fig2}(cc)). The peculiarity of this device is that both forward and backward coupled modes contribute to the NbN absorptance maximum. The absorptance maxima are attained inside a narrow plasmonic pass band, which originates from inversion of the plasmonic minigap at the top of the first Brillouin zone \cite{article49}. At small tilting the absorptance characteristics of NCDDAI$_{\lambda}$-A at 1550 nm resembles to the inversion of that in NCDAI$_{\lambda}$-A. In NCDDAI$_{\lambda}$-A the 93.00\% global absorptance maximum appears near perpendicular incidence at 0.69$^{\circ}$. The wavelength-scaled pattern couples into forward and backward propagating modes with slightly different 1030.75 nm and 1055.2 nm wavelength, respectively. The wavelength of both coupled modes indicates their plasmonic nature. The absorptance decreases rapidly, than reaches a 51.24\% local maximum at 69.5$^{\circ}$, which is significantly smaller than the PBA originating from array of NbN loaded nano-cavities.

The optimization of half-wavelength-scaled NCTAI$_{1/2*\lambda}$-A-SNSPD resulted in 600.00 nm optimal periodicity and 96.50 nm nano-cavity width, and 145.81 nm nano-cavity length (Figure~\ref{fig1}(da)). The small 38.10 nm and 260.83 nm widths of vertical gold segments neighboring the MIM nano-cavities at their anterior/exterior side make possible to reach very large absorptance in spite of narrow absorbing NbN stripes (see Appendix Figure~\ref{figA4}(ca)). The dispersion diagram of NCTAI$_{1/2*\lambda}$-A indicates that the inspected 1550 nm is inside a plasmonic pass band originating from backward propagating surface modes coupled in -1 order (Figure~\ref{fig2}(da)) \cite{article49, article50, article51, article52, article53}. However, the absorptance enhancement is more strongly tilting dependent compared to NCDDAI$_{1/2*\lambda}$-A and is significant only inside a narrower regime inside the first Brillouin zone in spite of sub-wavelength pitch \cite{article45, article46, article47, article48}. This indicates that NCTAI$_{1/2*\lambda}$-A devices can be used efficiently under optimized illumination conditions. The course of absorptance in NCTAI$_{1/2*\lambda}$-A resembles to the NCDDAI$_{1/2*\lambda}$-A, however exhibits a more pronounced polar angle dependence. Namely, the absorptance is 13.25\% at perpendicular incidence and increases rapidly with the polar angle until 49.00$^{\circ}$, where it reaches the 94.49\% global absorptance maximum. At this orientation the integrated plasmonic pattern couples into backward propagating surface modes in -1 order. Their short 1041.51 nm wavelength indicates that these are plasmonic modes. The fingerprint of the PBA corresponding to array of NbN loaded nano-cavities is not observable.

The optimization of three-quarter-wavelength-scaled NCTAI$_{3/4*\lambda}$-A-SNSPD resulted in 795.83 nm periodicity at the middle of the optimization interval, and 109.18 nm nano-cavity width close to the upper bound (Figure~\ref{fig1}(db)). Both the cavity width and the accompanying 176.78 nm length is commensurate with those in NCDDAI$_{3/4*\lambda}$-A, and their ratio is similarly increased with respect to NCTAI$_{1/2*\lambda}$-A. The 222.98 nm and 163.96 nm width of vertical gold segments at anterior and exterior sides of MIM nano-cavities is more commensurate, while their ratio is reversal with respect to those in NCTAI$_{1/2*\lambda}$-A (see Appendix Figure~\ref{figA4}(cb)). The dispersion graph of the optimized NCTAI$_{3/4*\lambda}$-A device indicates again a plasmonic pass band \cite{article49, article50, article51, article52, article53}, where the global absorptance maximum is achieved at transitional tilting in a wider spectral interval compared to NCTAI$_{1/2*\lambda}$-A (Figure~\ref{fig2}(db)). The NCTAI$_{3/4*\lambda}$-A more efficiently couples the incident light in -1 order into backward propagating surface modes. As a result, larger absorption is achieved inside an inverted plasmonic band-gap at the top of the first Brillouin zone. The NCTAI$_{3/4*\lambda}$-A device’s absorptance characteristics is similar to the corresponding NCDDAI$_{3/4*\lambda}$-A, however all absorptance values are slightly larger and the extrema are shifted to smaller polar angles. The NCTAI$_{3/4*\lambda}$-A device shows 75.37\% absorptance already at perpendicular incidence, than by increasing tilting the absorptance increases until 19.37$^{\circ}$ polar angle, where 94.95\% global maximum is achieved. At this tilting light coupling occurs in -1 order into backward propagating surface modes having a wavelength of 1056.99 nm, which exhibit plasmonic characteristics. The absorptance decreases monotonously by increasing tilting, i.e. PBA corresponding to array of NbN loaded nano-cavities is not observable. 

The optimization of wavelength-scaled NCTAI$_{\lambda}$-A-SNSPD resulted in the largest absorptance in 1056.24 nm pitch integrated pattern, which almost equals to the wavelength of plasmons propagating at silica-gold interface. The 103.69 nm optimal width of nano-cavities is the smallest among wavelength-scaled integrated devices (Figure~\ref{fig1}(dc)). The optimal 183.2 nm nano-cavity length is larger than the length in corresponding NCTAI$_{1/2*\lambda}$-A and NCTAI$_{3/4*\lambda}$-A devices, and it is decreased with respect to optimal nano-cavity length in NCDDAI$_{\lambda}$-A. The 219.46 nm and 219.03 nm widths of vertical gold segments at the anterior and exterior sides of the nano-cavities is almost the same (see Appendix Figure~\ref{figA4}(cc)). Among NCTAI-A devices the dispersion characteristics of NCTAI$_{\lambda}$-A indicates the smallest extension of frequency-polar angle region, where significant absorptance enhancement occurs (Figure~\ref{fig2}(dc)). Both forward and backward coupled modes contribute to the absorptance maximum at the zones’ crossing point. The largest absorptance is attained inside an inverted plasmonic minigap at the top of the first Brillouin zone similarly to NCDDAI$_{\lambda}$-A\cite{article49}, however the achieved absorptance is larger. The absorptance characteristics of NCTAI$_{\lambda}$-A is very similar to that of NCDDAI$_{\lambda}$-A. Most important result of this work is that NCTAI$_{\lambda}$-A shows the highest achievable NbN absorptance among all inspected SNSPD devices, namely 95.05\% absorptance is achieved at perpendicular incidence. The wavelength-scaled pattern couples into forward and backward propagating modes in 1 and -1 order with the same 1056.24 nm wavelength, which refers to SPPs propagating at the gold-substrate interface. The polar angle dependent absorptance decreases first dramatically, than reaches a 40.82\% local maximum at 58.8$^{\circ}$, which is smaller than the PBA corresponding to NbN loaded nano-cavities. 

\subsection{Role of Near-Field Distribution and Volume-Fraction-Ratio of Absorbing Segments} 
\label{near-field}

\begin{figure}[h]
\centerline{\includegraphics[width=0.8\columnwidth,draft=false]{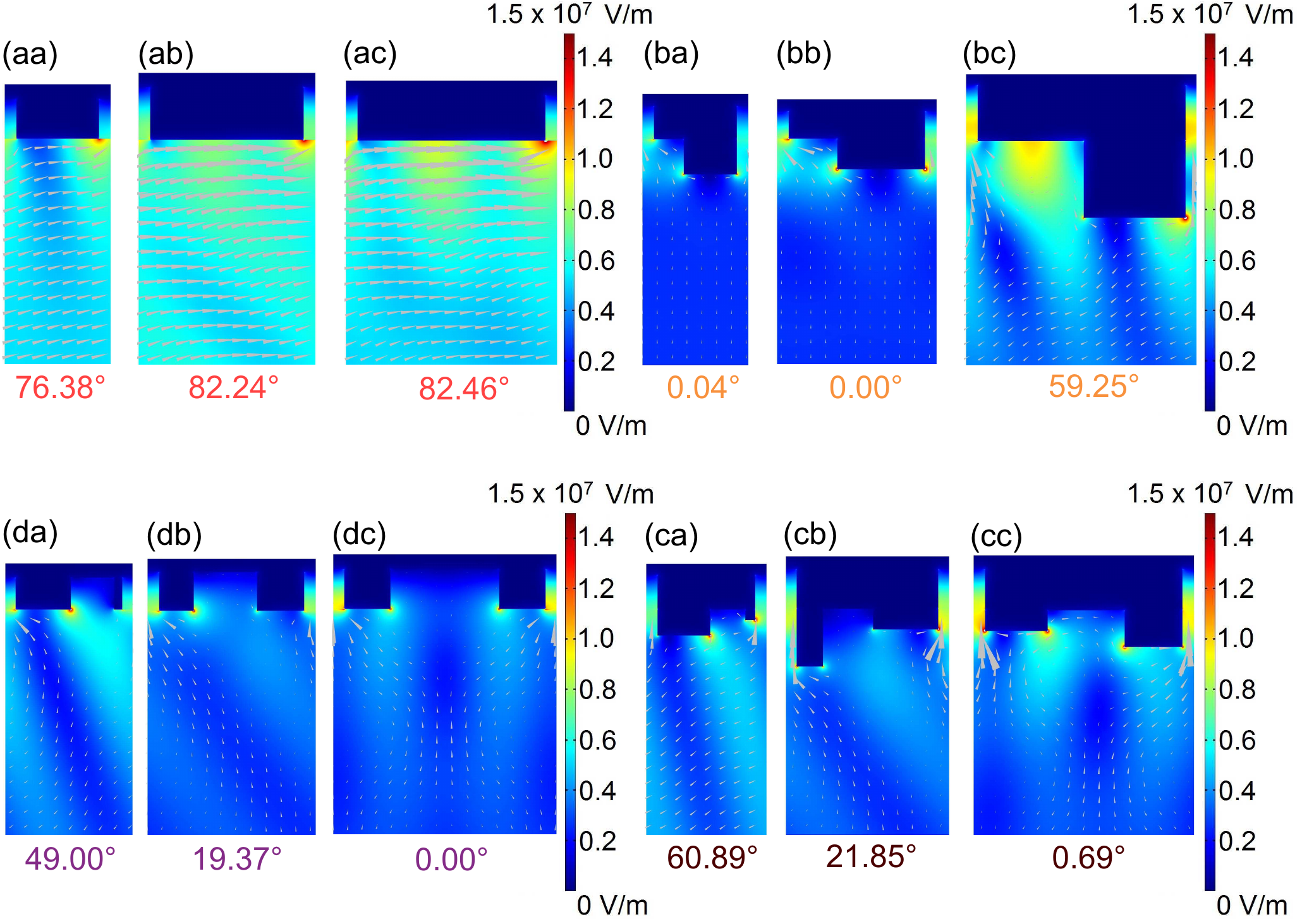}}
\caption{Time-averaged \textbf{E}-field and power flow at absorptance maxima in polar angle in optimized (aa) NCAI$_{1/2*\lambda}$-A, (ab) NCAI$_{3/4*\lambda}$-A, (ac) NCAI$_{\lambda}$-A, (ba) NCDAI$_{1/2*\lambda}$-A, (bb) NCDAI$_{3/4*\lambda}$-A, (bc) NCDAI$_{\lambda}$-A, (ca) NCDDAI$_{1/2*\lambda}$-A, (cb) NCDDAI$_{3/4*\lambda}$-A, (cc) NCDDAI$_{\lambda}$-A, (da) NCTAI$_{1/2*\lambda}$-A, (db) NCTAI$_{3/4*\lambda}$-A and (dc) NCTAI$_{\lambda}$-A.}
\label{fig3}
\end{figure}

The limits in achievable absorptance can be understood by comparing the \textbf{E}-field time-evolution in optimized devices and the volume-fraction-ratio of absorbing materials (Multimedia files 1-12, see Appendix Figure~\ref{figA1}-\ref{figA4}). The distribution of the time-averaged near-field at the maxima in each optimized NCAI-SNSPD type device confirms that the \textbf{E}-field is strongly enhanced at the entrance of NbN loaded MIM nano-cavities with $\sim$ $<(\lambda/4)$ length at tilting corresponding to PBA and the reflected wave’s intensity is negligible (Figure~\ref{fig3}(aa-ac), see Multimedia files 1-3). The \textbf{E}-field enhancement is strengthen by light tunneling through the MIM nano-cavities according to the PBA related nanophotonical phenomena \cite{article42, article43, article44}. The Poynting vector is almost parallel to the substrate-integrated structure interface at close proximity of the gold segments while it points to the cavities at their entrance.

In NCAI$_{1/2*\lambda}$-A the $0.71*(\lambda/4)$ nano-cavity length corresponds to strongly squeezed resonant MIM modes (see Appendix Figure~\ref{figA1}(ba)). The time-evolution shows that the \textbf{E}-field is highly enhanced in at least one of the neighboring MIM cavities, and both are shined efficiently in significant part of each duty-cycle. Among NCAI-A-SNSPD devices the optimized NCAI$_{1/2*\lambda}$-A possess the largest $4.07*10^{-3}$ NbN/Au volume-fraction-ratio (see Appendix Figure~\ref{figA1}(ca)). In the optimized NCAI$_{3/4*\lambda}$-A the longer $0.95*(\lambda/4)$ cavity reveals that the resonant MIM modes are less squeezed than in NCAI$_{1/2*\lambda}$-A, but are still capable of ensuring \textbf{E}-field enhancement at the entrance of nano-cavities (see Appendix Figure~\ref{figA1}(bb)). Although, the neighboring cavities are still shined alternately, a significant part of illuminating beam overlaps with the inserted gold segments. The optimized NCAI$_{3/4*\lambda}$-A exhibits intermediate $1.85*10^{-3}$ NbN/Au volume-fraction-ratio (see Appendix Figure~\ref{figA1}(cb)). In NCAI$_{\lambda}$-A the nano-cavity length commensurate with $0.81*(\lambda/4)$ reveals that the resonant MIM modes are squeezed at medium level (see Appendix Figure~\ref{figA1}(bc)). However, a completely distinct \textbf{E}-field time evolution is observable in neighboring cavities. Because of the largest periodicity, the illumination of MIM cavities occurs in the smallest fraction of each duty-cycle, while considerable enhancement is observable below the inserted gold segments as well. This explains that the NbN absorptance is smaller than in the two pervious devices, in spite of the largest \textbf{E}-field enhancement at the MIM nano-cavity entrances. The absorptance improving effect is the least efficient in the NCAI$_{\lambda}$-A device, according to the smallest $1.68*10^{-3}$ NbN/Au volume-fraction-ratio (see Appendix Figure~\ref{figA1}(cc)).

In optimized NCDAI$_{1/2*\lambda}$-A the cavity-segment closed by NbN is commensurate with $0.59*(\lambda/4)$, while the $1.22*(\lambda/4)$ extended cavity length is commensurate with, but slightly larger than quarter-wavelength, which can support $(\lambda/4)$-type resonant MIM modes (see Appendix Figure~\ref{figA2}(ba)). The time-averaged near-field reveals that the \textbf{E}-field is significantly enhanced at the entrance of nano-cavities as well as on the deflector corners (Figure~\ref{fig3}(ba)). The Poynting vector indicates a backward directed power-flow, proving that the gold deflector array re-directs the incident light towards the preceding nano-cavities. The \textbf{E}-field time-evolution shows no reflected waves, on the contrary the deflectors efficiently guide the nearly perpendicularly incident light towards the NbN segments (see Multimedia files 4). The nano-cavity entrances and deflector corners are shined alternately, and the delay introduced by deflectors into the guided wave propagation ensures that one of neighboring cavities is illuminated throughout dominant part of each duty-cycle. Compared to NCAI$_{1/2*\lambda}$-A larger absorptance is achievable in spite of smaller $3.29*10^{-3}$ NbN/Au volume-fraction-ratio (see Appendix Figure~\ref{figA2}(ca)), due to the larger time-averaged local \textbf{E}-field enhancement around the NbN segments.

In optimized NCDAI$_{3/4*\lambda}$-A the length of nano-cavity segments closed by NbN is commensurate with $0.49*(\lambda/4)$, i.e. it is two-times smaller compared to those in NCAI$_{3/4*\lambda}$-A. The $1.04*(\lambda/4)$ length of extended cavities is just slightly larger than quarter-wavelength, indicating that they can support $(\lambda/4)$-type resonant MIM modes (see Appendix Figure~\ref{figA2}(bb)). Both the time-averaged \textbf{E}-field and Poynting vector distribution as well as the \textbf{E}-field time-evolution is very similar to those in NCDAI$_{1/2*\lambda}$-A (Figure~\ref{fig3}(bb)). The \textbf{E}-field is strongly and almost synchronously enhanced at the entrance of neighboring nano-cavities, while at the deflector corners the \textbf{E}-field enhancement is not symmetrical and is out-of-phase (see Multimedia files 5). Synchronous illumination of neighboring cavities results in that they contribute to the \textbf{E}-field enhancement in significant fraction of each duty-cycle. The $2.35*10^{-3}$ NbN/Au volume-fraction-ratio is just slightly reduced compared to NCDAI$_{1/2*\lambda}$-A (see Appendix Figure~\ref{figA2}(cb)). The competitive gold absorption is compensated via increased periodicity and decreased deflector length, which promotes to attain relatively high absorptance. 

In optimized NCDAI$_{\lambda}$-A the nano-cavity segments closed by NbN have $(\lambda/4)$ length, while the $2.38*(\lambda/4)$ length of extended cavities overrides half-wavelength, indicating that they can support $\sim3*(\lambda/4)$ type resonant MIM modes (see Appendix Figure~\ref{figA2}(bc)). The \textbf{E}-field enhancement is still large at the entrance of nano-cavities and the Poynting vector points towards the NbN segments, however the \textbf{E}-field becomes more significant at the cavity-side corner of deflectors and under the inserted gold segment as well (Figure~\ref{fig3}(bc)). The \textbf{E}-field time-evolution shows that the preceding nano-cavities, inserted gold segments, deflectors and succeeding nano-cavities are illuminated successively (see Multimedia files 6). This proves that the backward propagating modes coupled in -2 order on the integrated structure promote efficient illumination through each duty-cycle. Two \textbf{E}-field antinodes are observable vertically along extended cavities and in each period, according to the order of grating couplings at play \cite{article47, article48}. However, the $8.89*10^{-4}$ NbN/Au volume-fraction-ratio is significantly decreased compared to NCDAI$_{1/2*\lambda}$-A and NCDAI$_{3/4*\lambda}$-A (see Appendix Figure~\ref{figA2}(cc)). In addition to this, weak reflected waves are also noticeable in the near-field of this device indicating that conversion into bound surface waves is less efficient. Accordingly, the absorptance improving effect of the integrated gold nano-cavity and deflector grating is the least efficient in the NCDAI$_{\lambda}$-A among NCDAI-A-SNSPD devices. 

In optimized NCDDAI$_{1/2*\lambda}$-A the nano-cavity-segment closed by NbN is commensurate with $0.62*(\lambda/4)$, while the $0.81*(\lambda/4)$/$1.09*(\lambda/4)$ extended cavity length is slightly smaller/larger than quarter-wavelength for the deflector at the anterior/exterior side (see Appendix Figure~\ref{figA3}(ba)). This indicates that symmetrical $(\lambda/4)$ type resonant MIM modes are supported in case of illumination at either side of extended cavities. The \textbf{E}-field is enhanced at the NbN segments and at the corners on the exterior side of both deflectors (Figure~\ref{fig3}(ca)). The Poynting vector also indicates a power-flow towards both nano-cavities. In between the two deflectors both the \textbf{E}-field and the power-flow is weak, the Poynting vector is parallel with the substrate-gold interface and is directed towards the smaller anterior-side deflector. The \textbf{E}-field time-evolution shows that the incident light is effectively directed towards the NbN segments by double deflectors (see Multimedia files 7). The double deflectors with different size result in efficient, however slightly asynchronous illumination of the MIM nano-cavities acting as $\sim(\lambda/4)$ resonators. Weak backward propagating surface modes, which originate from -1 order grating coupling, are also observable. The $3.08*10^{-3}$ NbN/Au volume fraction ratio is slightly smaller than in NCAI$_{1/2*\lambda}$-A and NCDAI$_{1/2*\lambda}$-A (see Appendix Figure~\ref{figA3}(ca)). Accordingly, the absorptance maximum is smaller in presence of double deflectors. 

In optimized NCDDAI$_{3/4*\lambda}$-A the nano-cavity segments closed by NbN is commensurate with $0.71*(\lambda/4)$, while the $1.08*(\lambda/4)$/$1.75*(\lambda/4)$ extended cavity lengths indicate that quarter/half-wavelength type resonant MIM modes are supported in case of illumination at the anterior/exterior deflector side (see Appendix Figure~\ref{figA3}(bb)). The \textbf{E}-field is enhanced around the NbN segments and at the cavity-side corner of those deflectors, which are positioned at the anterior-side of nano-cavities (Figure~\ref{fig3}(cb)). No significant \textbf{E}-field enhancement is observable below the inserted gold segment. The Poynting vectors indicate right/left directed power-flow towards the NbN segments along deflectors positioned at their anterior/exterior sides. The cavity resonances in $\sim(\lambda/4)$ and $\sim(\lambda/2)$ modes shine the NbN segments in neighboring nano-cavities alternately. The \textbf{E}-field time-evolution indicates well defined backward propagating waves originating from -1 order coupling, while very weak reflected waves are observable under the integrated pattern (see Multimedia files 8). The $1.74*10^{-3}$ NbN/Au volume fraction ratio is reduced with respect to NCDDAI$_{1/2*\lambda}$-A (see Appendix Figure~\ref{figA3}(cb)). Accordingly, the achieved maximal absorptance is smaller. 

In optimized NCDDAI$_{\lambda}$-A the length of cavity segment closed by NbN is commensurate with $0.74*(\lambda/4)$. Although, the $1.40*(\lambda/4)$ and $1.11*(\lambda/4)$ extended cavity lengths reveal asymmetry, $(\lambda/4)$ type resonant MIM modes are supported in either case of anterior and exterior side illumination (see Appendix Figure~\ref{figA3}(bc)). The \textbf{E}-field is enhanced at the entrance of nano-cavities and at cavity-side corners of both deflectors (Figure~\ref{fig3}(cc)). Caused by increased periodicity, double deflector array cannot expel the \textbf{E}-field from the region in between them, as a consequence clockwise power-flow vortices are noticeable under the inserted gold segments. The \textbf{E}-field time-evolution shows standing waves, however with time-dependent intensity (see Multimedia files 9). These standing waves originate from the co-existent forward and backward propagating plasmonic surface modes coupled in +1 and -1 order on the integrated pattern, which ensure that the neighboring nano-cavities are almost synchronously illuminated. The right phase-shift is introduced by the slightly different $1.40*(\lambda/4)$ and $1.11*(\lambda/4)$ extended nano-cavity lengths, and the time evolution of \textbf{E}-field compensates the asymmetry of deflectors. In the scattered-field region the decay of these standing modes rather than reflected light is observable. The $1.34*10^{-3}$ NbN/Au volume fraction ratio is further reduced, so the achievement of large NbN absorptance can be understood by taking the near-field distribution into account (see Appendix Figure~\ref{figA3}(cc)). According to slightly larger NbN/Au volume-fraction-ratio, larger absorptance is achievable than in NCDAI$_{\lambda}$-A.

In optimized NCTAI$_{1/2*\lambda}$-A both the nano-cavity width and length are decreased. The $0.57*(\lambda/4)$ length of nano-cavities is significantly smaller than quarter-wavelength, which reveals that the MIM modes are more strongly squeezed than in NCAI$_{1/2*\lambda}$-A (see Appendix Figure~\ref{figA4}(ba)). At the global absorptance maximum \textbf{E}-field enhancement is observable both at the anterior and exterior sides of the cavities, and at the anterior side of trenches (Figure~\ref{fig3}(da)). The Poynting vectors indicate power-flow towards the entrance of nano-cavities, and the time-averaged \textbf{E}-field is negligible inside the trench. The neighboring cavities and the inserted trench are lightened successively (see Multimedia files 10). The backward propagating waves are only weakly observable caused by co-existent intense reflected waves. The $4.53*10^{-3}$ NbN/Au volume fraction ratio is the largest among half-wavelength-scaled devices (see Appendix Figure~\ref{figA4}(ca)). In spite of larger NbN/Au volume-fraction-ratio, the \textbf{E}-field enhancement is slightly weaker and the resulted absorptance is slightly smaller, than in NCDDAI$_{1/2*\lambda}$-A. This can be explained by that one degree of freedom, namely that related to deflector's length difference, is not at play during NCTAI-A optimization.

In the optimized NCTAI$_{3/4*\lambda}$-A device the $0.69*(\lambda/4)$ cavity length is smaller than quarter wavelength, however the squeezing of localized modes resonant in the MIM nano-cavities is smaller with respect to NCTAI$_{1/2*\lambda}$-A (see Appendix Figure~\ref{figA4}(bb)). The \textbf{E}-field distribution is very similar to that in NCTAI$_{1/2*\lambda}$-A (Figure~\ref{fig3}(db)). The Poynting vector indicates a power-flow towards the nano-cavities, and the \textbf{E}-field is completely expelled from the region of trenches. The neighboring cavities are illuminated alternately, while the trench is lightened almost continuously (see Multimedia files 11). Well-defined backward propagating waves are observable, which originate from -1 order coupling. The $3.60*10^{-3}$ NbN/Au volume fraction ratio is the largest among three-quarter-wavelength scaled devices (see Appendix Figure~\ref{figA4} (cb)). The enhancement and the resulted absorptance is larger than in NCDDAI$_{3/4*\lambda}$-A, in accordance with two-times larger NbN/Au volume-fraction-ratio. 

In the optimized NCTAI$_{\lambda}$-A the $0.72*(\lambda/4)$ nano-cavity length indicates that the resonant modes are the less squeezed, when the periodicity of in-plane vertical gold segments is wavelength-scaled (see Appendix Figure~\ref{figA4} (bc)). The \textbf{E}-field is enhanced significantly at the entrances of nano-cavities, while at the trench corners only smaller enhancement is observable (Figure~\ref{fig3}(db)). The Poynting vector shows strong perpendicular power-flow towards the NbN segments. Standing waves are observable, which originate from the co-existent forward and backward propagating coupled modes, as a result the neighboring cavities and the inserted trenches are illuminated alternately. The \textbf{E}-field enhancement is significant around NbN segments in dominant parts of the duty-cycles. As a result, high absorptance enhancement is achieved in the NCTAI$_{\lambda}$-A, in accordance with the two-times larger $2.66*10^{-3}$ NbN/Au volume-fraction-ratio compared to NCDDAI$_{\lambda}$-A, which is the largest among wavelength-scaled devices (see Appendix Figure~\ref{figA4} (cc)). The absorptance enhancement is the highest in NCTAI$_{\lambda}$-A in spite of the fact that the NbN/Au volume fraction ratio is reduced compared to other NCTAI-A devices.

\subsection{Polarization Contrast Attainable via A-SNSPDs}
\label{polc}

\begin{figure}[h]
\centerline{\includegraphics[width=1\columnwidth,draft=false]{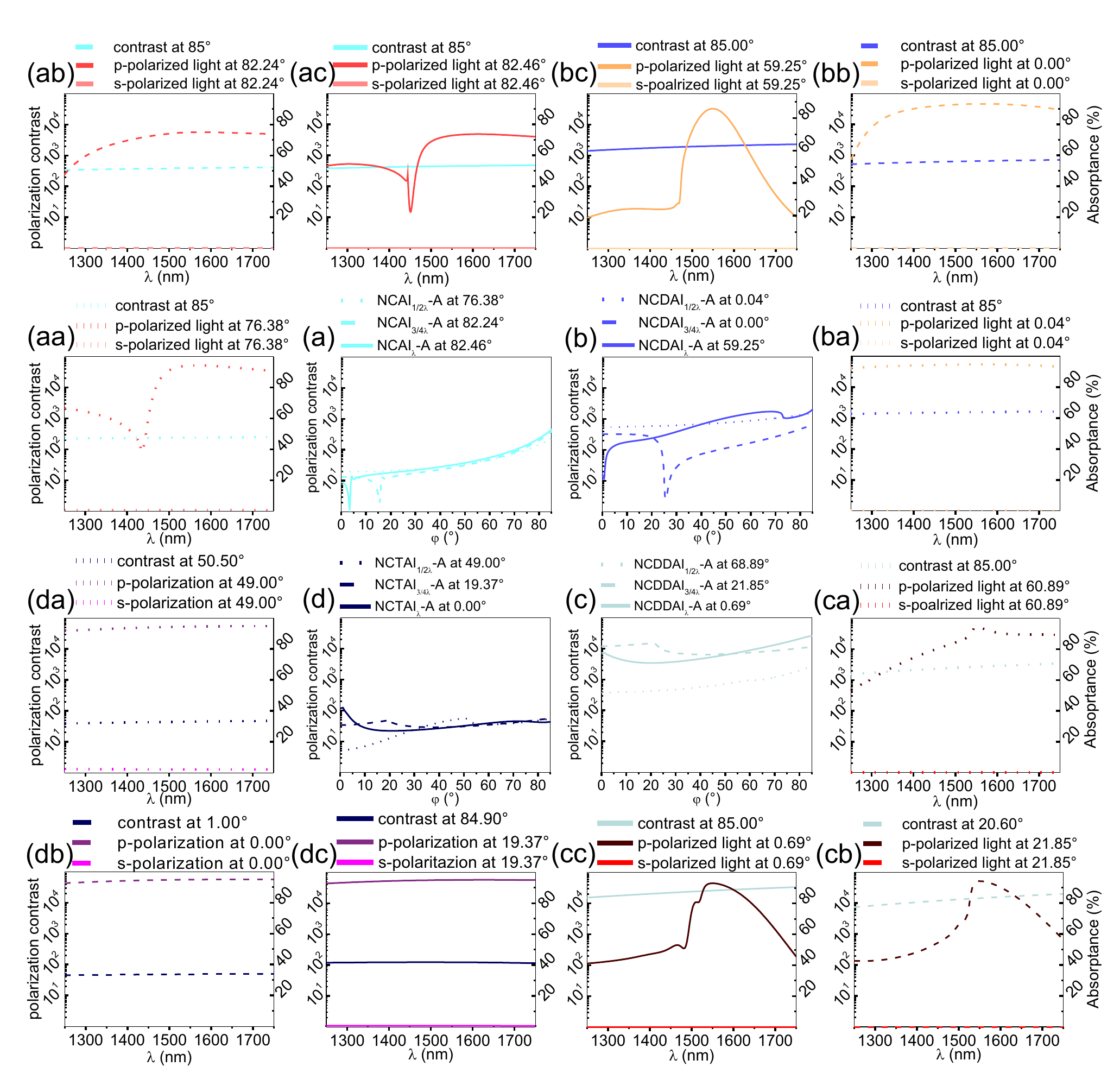}}
\caption{Polar angle dependent p-to-s polarization contrast achievable via (a) NCAI-A, (b) NCDAI-A, (c) NCDDAI-A and (d) NCTAI-A-SNSPD. Wavelength dependent p-to-s polarization contrast at contrast maxima and wavelength dependent NbN absorptance at absorptance maxima in (aa) NCAI$_{1/2*\lambda}$-A, (ab) NCAI$_{3/4*\lambda}$-A, (ac) NCAI$_{\lambda}$-A, (ba) NCDAI$_{1/2*\lambda}$-A, (bb) NCDAI$_{3/4*\lambda}$-A, (bc) NCDAI$_{\lambda}$-A, (ca) NCDDAI$_{1/2*\lambda}$-A, (cb) NCDDAI$_{3/4*\lambda}$-A, (cc) NCDDAI$_{\lambda}$-A, (da) NCTAI$_{1/2*\lambda}$-A, (db) NCTAI$_{3/4*\lambda}$-A and (dc) NCTAI$_{\lambda}$-A-SNSPD.}
\label{fig4}
\end{figure}

High polarization contrast is crucial in numerous quantum information processing applications \cite{article34, article35}. In A-SNSPD type devices the polarization contrast exponentially increases by increasing tilting almost throughout the complete polar angle interval, except the regions, where it is modulated dominantly by the polar angle dependent NbN absorptance of p-polarized light (Figure~\ref{fig4}(a-d)). One exception is the NCTAI$_{\lambda}$-A, where the polarization first decreases at small polar angle and after a global minimum increases exponentially. Figure~\ref{fig4}(a) indicates that the polarization contrast of NCAI$_{1/2*\lambda}$-A is 19.28 at perpendicular incidence and increases monotonically exponentially. The increase is the less significant in this device, and the polarization contrast reaches 240.92 at 85.00$^{\circ}$. On the polarization contrast of NCAI$_{3/4*\lambda}$-A a significant modulation appears at $\sim15.60^{\circ}$ tilting according to the global minimum in p-polarized light absorptance. At perpendicular incidence the polarization contrast is slightly smaller than in NCAI$_{1/2*\lambda}$-A, but increases with a larger rate through 389.45 at 85$^{\circ}$. The course of polarization contrast in NCAI$_{\lambda}$-A is similar to that of NCAI$_{3/4*\lambda}$-A, the only significant difference is the shift of modulation towards smaller polar angles. Although, the contrast at perpendicular incidence is smaller than in the two previous devices, it reaches 450.35 value at 85.00$^{\circ}$, which is the largest among NCAI-SNSPDs. The parallel exponential decrease of s-polarized absorptance in S-orientation suppresses the modulation originating from PBA on the p-polarized absorptance of NCAI-A-SNSPDs.

In deflector-array integrated NCDAI-SNSPDs the polarization contrast overrides those in counterpart NCAI-SNSPDs throughout dominant part of the inspected polar angle interval, which proves the polarization selective role of deflectors (Figure~\ref{fig4}(b)). In NCDAI$_{1/2*\lambda}$-A the polarization contrast is an order of a magnitude higher than in the counterpart NCAI$_{1/2*\lambda}$-A device. At perpendicular incidence the contrast is 552.68 and increases exponentially through 1552.96 at 85.00$^{\circ}$, without any significant modulation. In NCDAI$_{3/4*\lambda}$-A the contrast is $\sim300$ in the 0.00$^{\circ}$ - 20.00$^{\circ}$ polar angle interval, according to the plateau in polar angle dependent absorptances. At $\sim25.50^{\circ}$ orientation corresponding to the cut-off in p-polarized absorptance a global minimum arises, than the contrast increases and reaches 656.51 at 85.00$^{\circ}$. In NCDAI$_{\lambda}$-A the polarization contrast is 15.86 at perpendicular incidence and rapidly increases in a 2$^{\circ}$ interval according to the same characteristics of the p-polarized absorptance. By increasing tilting the contrast increases through the 2029.44 global maximum at 85.00$^{\circ}$.

In double-deflector integrated NCDDAI-SNSPDs the polarization contrast is further enhanced (Figure~\ref{fig4}(c)). In NCDDAI$_{1/2*\lambda}$-A at perpendicular incidence the 381.07 contrast is less than in NCDAI$_{1/2*\lambda}$-A, but increases monotonically exponentially trough 2678.96 at 85.00$^{\circ}$. In NCDDAI$_{3/4*\lambda}$-A significantly larger polarization contrast is reachable. The contrast is 11915.59 already at perpendicular incidence and increases monotonically until 20.60$^{\circ}$, where it reaches a global maximum of 14498.44. After this global maximum a small reduction occurs originating from the course of polar angle dependent p-polarized absorptance, than the contrast increases slightly according to more rapid exponential decrease of s-polarized absorptance. In NCDDAI$_{\lambda}$-A the contrast is 7254.79 already at perpendicular incidence, than decreases according to the relative slopes of p- and s-polarized absorptance decreases. After a global minimum of 3426.28 at 20.1$^{\circ}$ the contrast increases through 25853.17 at 85.00$^{\circ}$.

The trench array integrated NCTAI-SNSPDs result in the smallest polarization contrast indicating that embedded trenches enhance the absorptance via less polarization selective phenomena (Figure~\ref{fig4}(d)). In NCTAI$_{1/2*\lambda}$-A the contrast is 5.16 at perpendicular incidence and reaches its global maximum of 55.31 at 50.50$^{\circ}$ according to the large maximum on the polar angle dependent p-polarized absorptance. In NCTAI$_{3/4*\lambda}$-A the contrast is 33.88 at perpendicular incidence and increases monotonically until 19.9$^{\circ}$, where a global/local maximum is observable in absorptance/contrast. Finally the contrast reaches its maximum of 54.63 at 84.90$^{\circ}$. In NCTAI$_{\lambda}$-A the contrast shows a global maximum of 123.75 at 1.00$^{\circ}$, than begins to decrease rapidly according to the polar angle dependent absorptance. After a global minimum at 20.9$^{\circ}$ the contrast reaches 45.19 local maximum at 72.00$^{\circ}$.  

\subsection{Bandwidth Corresponding to Maxima in Absorptance and in Polarization Contrast Achievable via Optimized A-SNSPD Configurations}
\label{bandwidth}

On the wavelength dependent p-polarized absorptance at tilting corresponding the plasmonic Brewster angle of NCAI-A devices a cut-off is observable in the inspected region. The cut-off in NCAI$_{1/2*\lambda}$-A corresponds to the first Wood-anomaly $(d(\sin\varphi+\sin90^{\circ})=1*\lambda)$, in accordance with the literature (Figure~\ref{fig4}(aa)) \cite{article63}. Our present work demonstrates that in NCAI$_{\lambda}$-A device having two times larger periodicity a similar cut-off originating from second order Wood-anomaly $(2*d(\sin\varphi+\sin90^{\circ})=2*\lambda)$ arises (Figure~\ref{fig4}(a)/(a-c)). At tilting corresponding to maxima in polar angle the NbN absorptance is almost wavelength independent in NCDAI$_{1/2*\lambda}$-A, a cut-off is observable at the small wavelength edge of the inspected interval in NCDAI$_{3/4*\lambda}$-A, while the NCDAI$_{\lambda}$-A has a discrete plasmonic pass band revealing to polarization specific filtering capability \cite{article64} (Figure~\ref{fig4}(b)/(a-c)). 

In all NCDDAI-SNSPD devices the wavelength dependent absorptance indicates plasmonic pass band at tilting corresponding to maxima in polar angle, which is the broadest in NCDDAI$_{1/2*\lambda}$-A, asymmetrical in NCDDAI$_{3/4*\lambda}$-A, and almost symmetrical in NCDDAI$_{\lambda}$-A, revealing that different phenomena result in polarization specific filtering capability inside bands of different width (Figure~\ref{fig4}(c)/(a-c)) \cite{article64}. The absorptance at tilting corresponding to maxima in polar angle does not show wavelength dependency in any NCTAI-SNSPD, indicating that these devices are almost perfect absorbers in very broadband of several hundred nanometers (Figure~\ref{fig4}(d)/(a-c)). Moreover, the polarization contrast at tilting corresponding to maxima in polar angle does not indicate wavelength dependency in any SNSPD-A devices, i.e. commensurate polarization selectivity is preserved trough wide bands.

\section{Discussion and Conclusion}
\label{discuss}

In case of the simplest integrated NCAI-A-SNSPD type nano-cavity-grating profiles the collective resonance on the MIM nano-cavity-array enhanced by light tunneling at the PBA is capable of ensuring large absorptance enhancement at well-defined polar-angle in wide wavelength interval (Figure~\ref{fig1}(a),~\ref{fig2}(a),~\ref{fig3}(a),~\ref{fig4}(a)). The locations of maxima correspond to $\cos\theta_{Brewster}=\beta w/k_{0}p$ tilting, where $k_0$ and $\beta$ are the propagation constant of photonic and MIM modes, w is  nano-cavities' width, and p is the periodicity \cite{article42, article43, article44}. The absorptance is almost polar angle independent in NCAI$_{1/2*\lambda}$-A according to the sub-wavelength pitch \cite{article45, article46, article47, article48}, while grating coupling related modulations resulting in Wood-anomaly features perturb the absorptance in NCAI$_{3/4*\lambda}$-A and NCAI$_{\lambda}$-A \cite{article31, article33, article49, article50, article51, article52, article53, article54, article55, article56}. By increasing the periodicity of NCAI-A devices the NbN/Au volume fraction ratio gradually decreases (see Appendix Figure~\ref{figA1}(c)/(a-c)). The achieved absorptance rapidly decreases, while the similar optimal orientations are in accordance with the pitch dependence of PBA (see Appendix Figure~\ref{figA1} (a)/(a-c)). The maximal absorptance achieved at 1550 nm in NCAI-SNSPD devices is determined by the \textbf{E}-field enhancement attainable at the PBA corresponding to NbN loaded nano-cavities. The optimal integrated device geometry is NCAI$_{1/2*\lambda}$-A based on the highest attainable 94.18\% absorptance via 500 nm periodic NbN pattern at 76.38$^{\circ}$ tilting. The observed wavelength dependency of the achieved large absorptance at maxima arising at PBA indicates that it is possible to set the cut-off outside the region of interest via appropriate design \cite{article63}. The absorptance might be further enhanced not only in wide band but also in wide angle interval by applying tilted cavity walls and conical MIM cavity shapes in NCAI-A-SNSPDs \cite{article63, article65}. However, inspection of tilted walls was outside the scope of present optimization studies.

In case of NCDAI$_{1/2*\lambda}$-A the collective resonance is capable of ensuring large absorptance throughout the largest polar angle and wavelength interval \cite{article45, article46, article47, article48}. (Figure~\ref{fig2}(ba)). In NCDAI$_{3/4*\lambda}$-A the grating coupling in -1 order into Brewster-Zenneck type waves results in a cut-off (Figure~\ref{fig2}(bb)) \cite{ article49, article50, article51, article52, article53, article57, article58, article59, article60, article61}. In NCDAI$_{\lambda}$-A grating coupling in -2 order into backward propagating modes enhances the absorptance well before the second Brillouin zone boundary (Figure~\ref{fig2}(bc)). The surprisingly small 975.45 nm coupled mode wavelength is determined by the interplay of a plasmonic band and the PBA phenomenon corresponding to the large cavity arising between successive deflectors. The 55.55$^{\circ}$ PBA originating from deflector-array approximates the 59.25$^{\circ}$ tilting corresponding to the NbN absorptance maximum. The NbN/Au volume fraction ratio more rapidly decreases with the periodicity in presence of deflectors (see Appendix Figure~\ref{figA2}(c)/(a-c)). Accordingly, the achieved absorptance gradually decreases, while the significantly different optimal orientations reveal that various nanophotonical phenomena are at play in different periodicity intervals (see Appendix Figure~\ref{figA2} (a)/(a-c)). The optimal integrated device geometry is NCDAI$_{1/2*\lambda}$-A based on the highest attainable 94.68\% absorptance via 500.09 nm periodic integrated pattern at 0.04$^{\circ}$ tilting. The absorptance achieved via NCDAI$_{1/2*\lambda}$-A is larger than in NCAI$_{1/2*\lambda}$-A in spite of gold deflectors presence, further important advantage is that the global maximum appears at almost perpendicular incidence. 

All NCDAI-A devices exhibit well defined polarization dependent filtering capability through wide bands, which can be tailored by the integrated device design \cite{article64}. At tilting corresponding to maximal absorptance of NCDAI$_{\lambda}$-A the cut-off at small wavelength is coincident with the second Wood anomaly, while the bandwidth is small in spite of the nearby PBA originating from cavities in between deflectors. This suggests that better absorptance can be achieved, only when the grating coupling related maximum is coincident with PBA corresponding to NbN loaded nano-cavity-array. Under these circumstances besides co-existent grating coupling and light tunneling the wavelength independence of PBA can be preserved.

In case of NCDDAI-A-SNSPD type devices the coupling into plasmonic surface modes results in commensurately large global maxima, however at significantly different tilting (Figure~\ref{fig1}(c)). In half- and quarter-wavelength-scaled integrated devices the backward propagating surface waves coupled in -1 order are capable of resulting in enhanced absorptance. Moreover, strong-coupling between collectively resonating nano-cavity and surface modes is also observable in NCDDAI$_{1/2*\lambda}$-A, which is capable of resulting in larger absorptance enhancement, than that originating from cavity or surface mode related resonances separately \cite{article62, article66}. The global maximum appears at 60.89$^{\circ}$, which is close to 55.39$^{\circ}$ tilting, where the plasmonic band overlaps with PBA phenomenon corresponding to array of cavities arising between the longer exterior-side deflectors. Similarly, in NCDDAI$_{3/4*\lambda}$-A a plasmonic pass band determined maximum appears at 21.85$^{\circ}$ tilting, which is just slightly smaller than the 32.01$^{\circ}$ PBA originating from the cavity arising between the longer exterior-side deflectors.

 In NCDDAI$_{\lambda}$-A grating couplings in 1 and -1 order into forward and backward propagating modes enhances the absorptance at the top of the first Brillouin zone inside inverted minigaps \cite{article49}. By increasing the periodicity of NCDDAI-A devices the NbN/Au volume fraction ratio gradually decreases (see Appendix Figure~\ref{figA3}(c)/(a-c)). The achieved absorptance slowly, while the corresponding polar angle very rapidly decreases (see Appendix Figure~\ref{figA3}(a)/(a-c)). This indicates, that different grating coupling phenomena act in different periodicity regions. The optimal integrated device geometry is half-wavelength-scaled also in case of NCDDAI-SNSPD. The highest 94.60\% absorptance attained via 500.09 nm periodic NbN pattern at 60.89$^{\circ}$. Although, the 94.34\% global maximum does not override the 94.60\% attainable via NCDDAI$_{1/2*\lambda}$-A, the NCDDAI$_{3/4*\lambda}$-A device has a smaller kinetic inductance and so faster reset time. However, NCDDAI$_{\lambda}$-A devices resulting in slightly smaller absorptance at almost perpendicular incidence and ensuring competitive kinetic inductance are also promising for special application purposes. All NCDDAI-A exhibit well defined polarization dependent filtering capability inside finite bands, which width strongly depends on the periodicity \cite{article64}. In all NCDDAI-A at tilting corresponding to maximal absorptance the cut-off at small wavelength is coincident with the first Wood anomaly. However, the bandwidth is again small in spite of nearby PBAs originating from cavities arising in between the longer exterior-side deflectors in NCDDAI$_{1/2*\lambda}$-A and NCDDAI$_{3/4*\lambda}$-A. Flat bands can be tailored by ensuring better overlap between plasmonic bands and PBA phenomenon corresponding to NbN loaded nano-cavity-array also in NCDDAI-A-SNSPD devices. Inspection of tilted cavity and deflector walls and conical extended cavities is subject of further studies \cite{article63, article65}. The NCDAI-A and NCDDAI-A devices can be considered as multi-component diffraction gratings, which can have unique plasmonic band structure and can exhibit strong absorption at specific wavelength in case of optimized design \cite{article67, article68, article69}. Further research is in progress to replace deflectors, which fabrication is challenging, with in-plane multi-component gratings. 

In case of NCTAI-SNSPD type grating profiles the coupling on the periodic pattern results in appearance of plasmonic pass-bands and in large global maxima, however at different tilting (Figure~\ref{fig1}(d) - to – (c)). Both in NCTAI$_{1/2*\lambda}$-A - and NCTAI$_{3/4*\lambda}$-A the backward propagating surface waves coupled in -1 order are capable of resulting in enhanced absorptance. In NCTAI$_{1/2*\lambda}$-A the absorptance maximum at 49.00$^{\circ}$ is close to 55.15$^{\circ}$ tilting, where the plasmonic band overlaps with PBA phenomenon corresponding to array of nano-cavities arising between the wider exterior deflectors. In NCTAI$_{\lambda}$-A grating couplings in 1 and -1 order into forward and backward propagating modes at the top of the first Brillouin zone result in the highest absorptance inside an inverted minigap \cite{article49}. By increasing the periodicity of NCTAI-A devices the NbN/Au volume fraction ratio slowly decreases (see Appendix Figure~\ref{figA4}(c)/(a-c)). In contrast, the absorptance slowly increases, while the optimal tilting decreases similarly to NCDDAI (see Appendix Figure~\ref{figA4}(a)/(a-c)). The optimal integrated device geometry is NCTAI$_{\lambda}$-A, and the 95.05\% absorptance attained via 1042.83 nm periodic NbN pattern at 0.00$^{\circ}$ is the highest absorptance achieved in optimized SNSPDs. The achievement of absorptance maximum under these condition is in contradiction with the literature, describing that when the Rayleigh condition is met, a minimum in transmission and in corresponding absorption is expected \cite{article53, article54, article55, article56}. The considerably large NbN/Au volume fraction ratio makes possible to achieve very high, almost perfect absorptance in NCTAI$_{\lambda}$-A-SNSPD (see Appendix Figure~\ref{figA4}(c)/(a-c)). Important advantage of the optimized NCTAI$_{\lambda}$-A that the kinetic inductance of the detector device is low which results in fast reset time. Further important advantage is that the global maximum appears at perpendicular incidence. 

All NCTAI-A devices are peculiar, since the achieved absorptance increases by increasing periodicity in spite of decreasing NbN/gold volume fraction ratio, moreover the absorptance is wavelength independent through a very wide band, even though PBA phenomenon corresponding to trenches co-exist only in NCTAI$_{1/2*\lambda}$-A. The NCTAI-A devices can be considered as the simplest compound gratings \cite{article70, article71}. The effect of distributed gratings consisting of different periodic components as well as the possible advantages of tilted cavity and trench walls is subject of further studies \cite{article63, article65}.

In summary, our present work proves that maximization of absorptance in SNSPDs can be realized by optimizing the dispersion characteristics of plasmonic structure integrated devices. For QIP applications requirements are to increase secure communication rate of information encoded into polarization via polarization specific detection efficiency, and to promote high communication rate at specific polarization via decreased reset time. Based on our present studies performances of these parameters are reasonably compromised via NCDAI$_{\lambda}$-A and NCDDAI$_{\lambda}$-A-SNSPDs, which ensure high absorptance at specific polarization along with large contrast and short reset time due to the reduced length. In A-SNSPD devices the polarization contrast exponentially increases with the tilting and reaches its maximum at 85.00$^{\circ}$ in most of the inspected device types, exceptions are the NCDDAI$_{3/4*\lambda}$-A and the NCTAI-A-SNSPDs. The wavelength independency of the polarization contrast is advantageous, however the large tilting necessary to reach maximal polarization contrast is a drawback. When higher polarization contrast is required, C-SNSPD device configurations optimized with conditional absorptance offer a reasonable tradeoff. These devices are presented in the Appendix. Analyses of nanophotonical phenomena in P-SNSPD devices, which are optimized to ensure maximal polarization contrast, is in progress.  

\ack

This work was supported by the European Union and the European Social Fund through project “ELITeam - Establishment of the ELI Institute at the University of Szeged: foundation of interdisciplinary research in the field of lasers and their applications” (Project number: T\'{A}MOP-4.2.2.D-15/1/KONV-2015-0024). M\'{a}ria Csete acknowledges that the project was supported by the J\'{a}nos Bolyai Research Scholarship of the Hungarian Academy of Sciences. G\'{a}bor Szab\'{o} acknowledges the support of Hungarian Academy of Sciences. The authors would like to thank D\'{a}vid Mar\'{a}czi for figures preparation.

\appendixx{}

To provide an acceptable compromise between the achievable absorptance and polarization contrast, we have specified criteria regarding the minimal absorptance, and inspected the achievable polarization contrast as well as the parameters of optimized C-SNSPD configurations, by gradually decreasing the conditional absorptance. The criterion regarding the absorptance, that have to be parallel met, was varied with 0.25\% steps in 3\% interval of the maximal absorptance previously determined for each specific A-SNSPD device types. The related nonlinearly constrained optimization problems were solved by penalty function approach, the obtained results clearly indicate that the constraints were satisfied \cite{article72}.

\begin{figure}[h]
\centerline{\includegraphics[width=0.8\columnwidth,draft=false]{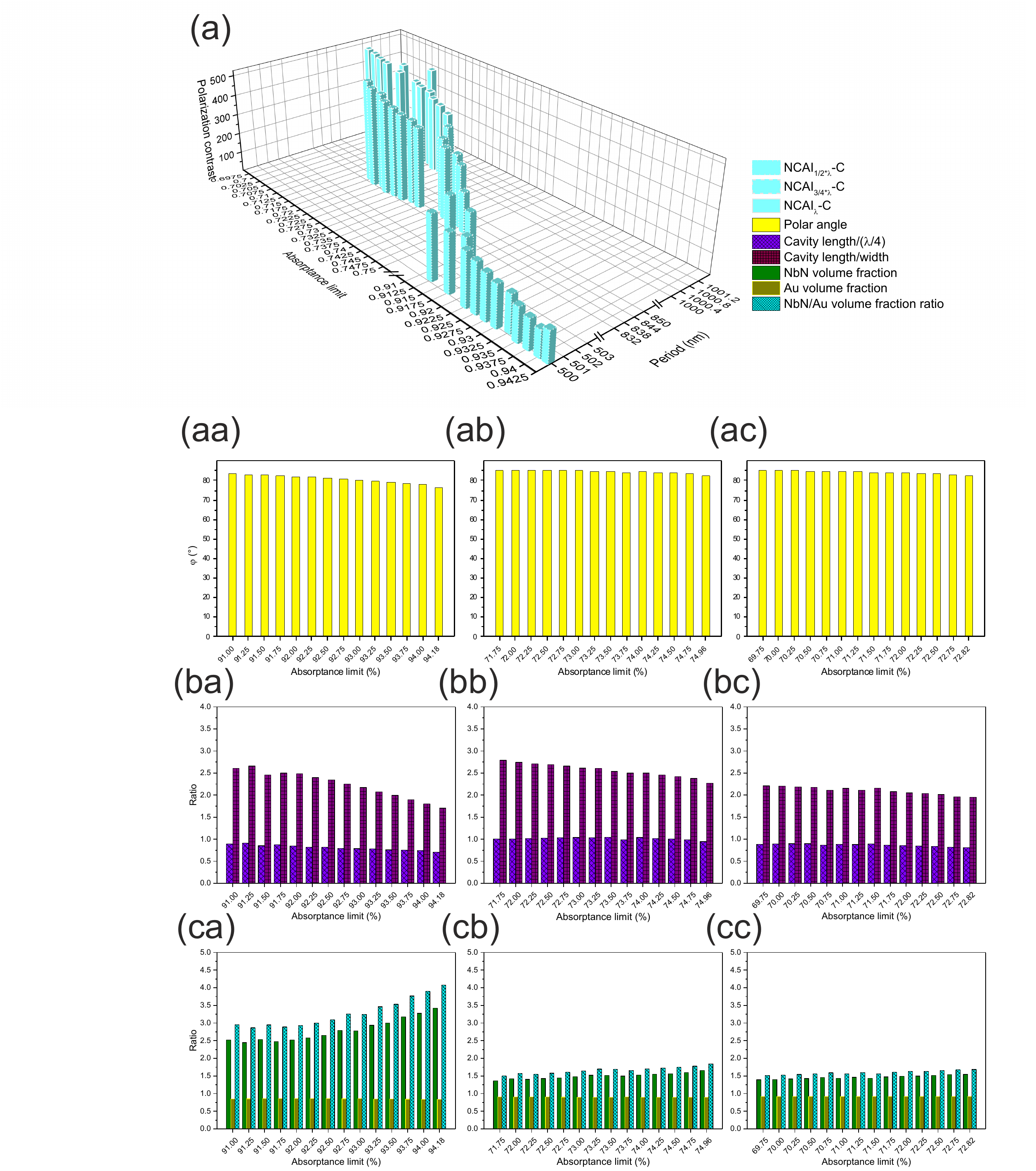}}
\caption{(a) Attained polarization contrast, optimized period and optimal polar angle as a function of conditional absorptance. Histograms indicating tendencies of (b) cavity length/$(\lambda/4)$ and cavity length/width ratios qualifying MIM modes squeezing, (c) NbN and Au volume fraction, NbN/Au volume fraction ratio, as a function of conditional absorptance, all in (a-c/a) NCAI$_{1/2*\lambda}$-C, (a-c/b) NCAI$_{3/4*\lambda}$-C, (a-c/c) NCAI$_{\lambda}$-C series.}
\label{figA1}
\end{figure}

\subsection{NCAI-C-SNSPDs Capable of Maximizing the Polarization Contrast at Conditional Absorptance}
In all NCAI-C-SNSPD devices the polarization contrast slowly and monotonously increases by decreasing the conditional absorptance (Figure~\ref{figA1}(a)). The polarization contrast increases in $[1.26*10^2-3.20*10^2]/[3.08*10^2-5.12*10^2]/[3.32*10^2-4.88*10^2]$ intervals in NCAI$_{1/2*\lambda}$-C/NCAI$_{3/4*\lambda}$-C/NCAI$_{\lambda}$-C. The period slightly modifies in [500.00 nm-502.61 nm]/[836.18 nm-835.09 nm]/[1000.00 nm-1000.02 nm] regions. All optimal polar angles correspond to the PBA, by decreasing the conditional absorptance their value increases slowly in [76.38$^{\circ}$-83.42$^{\circ}$]/[82.24$^{\circ}$-85.00$^{\circ}$]/[82.46$^{\circ}$-85.00$^{\circ}$] intervals, respectively (Figure~\ref{figA1}(a)/(a-c)). The cavity lengths non-monotonously increase in $[0.71*(\lambda/4)-0.89(\lambda/4)]/ [0.95*(\lambda/4)-1.01*(\lambda/4)]$ and $[0.81*(\lambda/4)-0.88*(\lambda/4)]$ intervals in NCAI$_{1/2*\lambda}$-C/NCAI$_{3/4*\lambda}$-C and NCAI$_{\lambda}$-C.

The cavity parameter tendencies result in monotonously increasing cavity length/width ratio, which reveals that the MIM modes become less confined in case of smaller conditional absorptance (Figure~\ref{figA1}(b)/(a-c)). The NbN/Au volume fraction ratio almost monotonously decreases in $[4.07*10^{-3}-2.96*10^{-3}]/[1.85*10^{-3}-1.50*10^{-3}]/[1.68*10^{-3}-1.52*10^{-3}]$ intervals (Figure~\ref{figA1}(c)/(a-c)).  

\subsection{NCDAI-C-SNSPDs Capable of Maximizing the Polarization Contrast at Conditional Absorptance}

\begin{figure}[h]
\centerline{\includegraphics[width=0.8\columnwidth,draft=false]{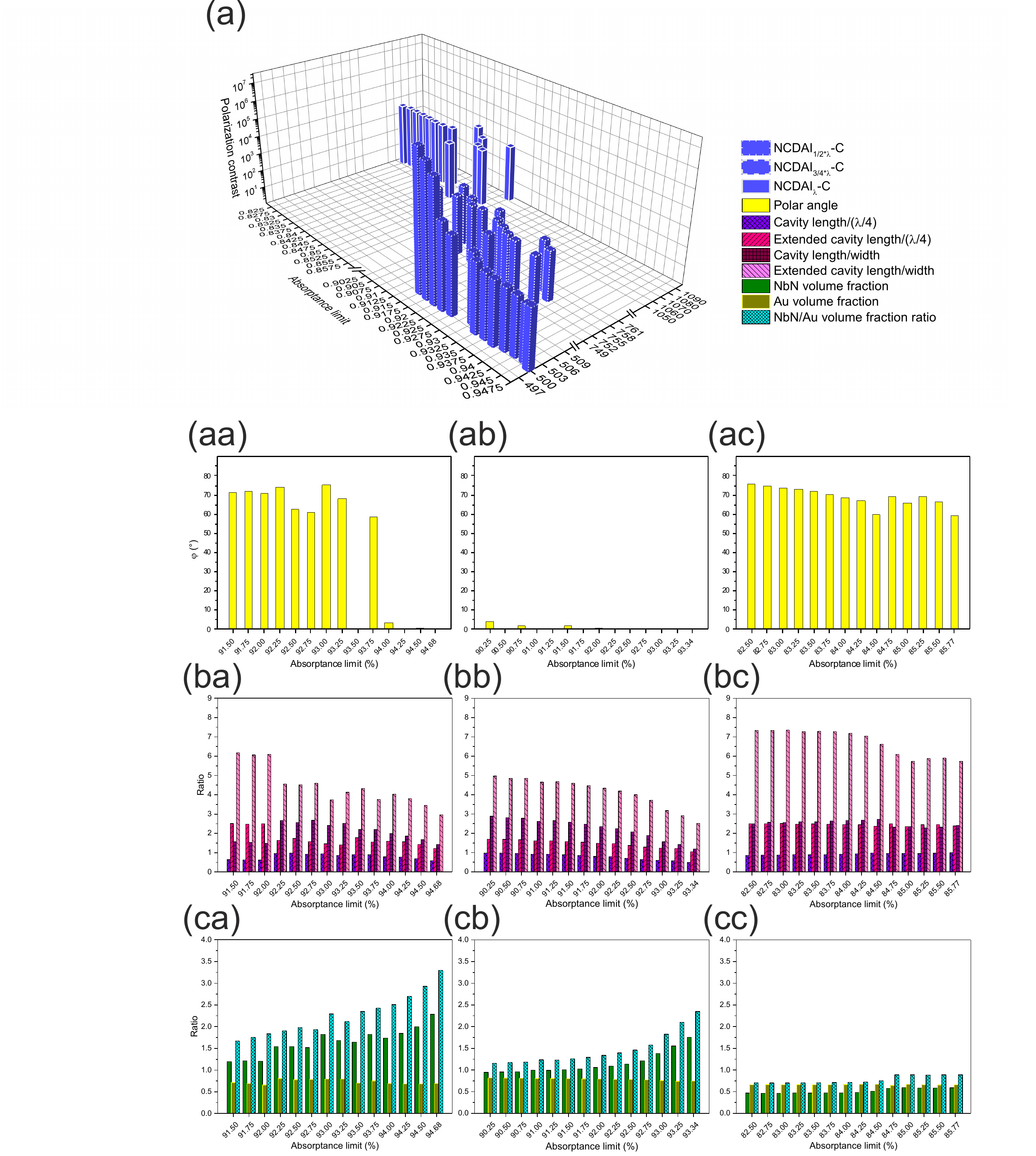}}
\caption{(a) Attained polarization contrast, optimized period and optimal polar angle as a function of conditional absorptance. Histograms indicating tendencies of (b) cavity length/$(\lambda/4)$, extended cavity length/$(\lambda/4)$, cavity length/width and extended cavity length/width ratios qualifying MIM modes squeezing, (c) NbN and Au volume fraction, NbN/Au volume fraction ratio as a function of conditional absorptance, all in (a-c/a) NCDAI$_{1/2*\lambda}$-C, (a-c/b) NCDAI$_{3/4*\lambda}$-C, (a-c/c) NCDAI$_{\lambda}$-C series.}
\label{figA2}
\end{figure}

In each NCDAI-C-SNSPD device the polarization contrast more rapidly and monotonously increases by decreasing the conditional absorptance (Figure~\ref{figA2}(a)). The polarization contrast increases in $[8.28*10^2-3.02*10^7]/[3.96*10^2-1.90*10^3]/[1.75*10^3-5.03*10^3]$ intervals in NCDAI$_{1/2*\lambda}$-C/NCDAI$_{3/4*\lambda}$-C/NCDAI$_{\lambda}$-C. The period non-monotonously varies in [500.09 nm-500.24 nm]/[754.61 nm-750.35 nm]/[1093.02 nm-1064.08 nm] regions. In NCDAI$_{1/2*\lambda}$-C the optimal polar angle jumps in [0.04$^{\circ}$-71.33$^{\circ}$] tilting region, indicating that different nanophotonical phenomena become to play to ensure conditional absorptance below a certain conditional absorptance level (Figure~\ref{figA2}(a)/(a-c)). In NCDAI$_{3/4*\lambda}$-C approximately perpendicular, [0$^{\circ}$-4.26$^{\circ}$] angle of incidence is optimal independently on the conditional absorptance, while in NCDAI$_{\lambda}$-C tilting in [59.25$^{\circ}$-75.65$^{\circ}$] region is preferred.

The cavity length exhibits a maximum in $[0.59*(\lambda/4)-0.64*(\lambda/4)]$ in NCDAI$_{1/2*\lambda}$-C/increases monotonously in $[0.49*(\lambda/4)-0.98*(\lambda/4)]$ interval in NCDAI$_{3/4*\lambda}$-C, while decreases in $[1.00*(\lambda/4)-0.85*(\lambda/4)]$ interval in NCDAI$_{\lambda}$-C (Figure~\ref{figA2}(b)/(a-c)). The cavity parameter tendencies result in cavity length/width ratio, which exhibits a maximum in NCDAI$_{1/2*\lambda}$-C and NCDAI$_{\lambda}$-C, and shows a monotonous increase in NCDAI$_{3/4*\lambda}$-C, respectively. The extended cavity length exhibits non-monotonous increase in $[1.22-2.51]*(\lambda/4)$ region in NCDAI$_{1/2*\lambda}$-C, and in $[2.38*(\lambda/4)-2.50*(\lambda/4)]$ region in NCDAI$_{\lambda}$-C, while slow monotonous increase can be seen in the region of $[1.04*(\lambda/4)-1.69*(\lambda/4)]$ in NCAI$_{3/4*\lambda}$-C. The deflector's presence manifests itself in non-monotonous/almost monotonous increase in extended cavity length/width ratio in NCDAI$_{1/2*\lambda}$-C/NCDAI$_{3/4*\lambda}$-C and NCDAI$_{\lambda}$-C. The type of MIM modes that are supported by extended cavities modifies from quarter- to three-times quarter-wavelength in NCDAI$_{1/2*\lambda}$-C, from quarter- to half-wavelength in NCDAI$_{3/4*\lambda}$-C, while remains three-quarter-wavelength in NCDAI$_{\lambda}$-C. The NbN/Au volume fraction ratio almost monotonously decreases in $[3.29*10^{-3}-1.67*10^{-3}]/[2.35*10^{-3}-1.16*10^{-3}][8.89*10^{-4}-7.07*10^{-4}]$ intervals in NCDAI$_{1/2*\lambda}$-C/NCDAI$_{3/4*\lambda}$-C/NCDAI$_{\lambda}$-C (Figure~\ref{figA2}(c)/(a-c)).
\newpage

\subsection{NCDDAI-C-SNSPD Capable of Maximizing the Polarization Contrast at Conditional Absorptance}

\begin{figure}[h]
\centerline{\includegraphics[width=0.8\columnwidth,draft=false]{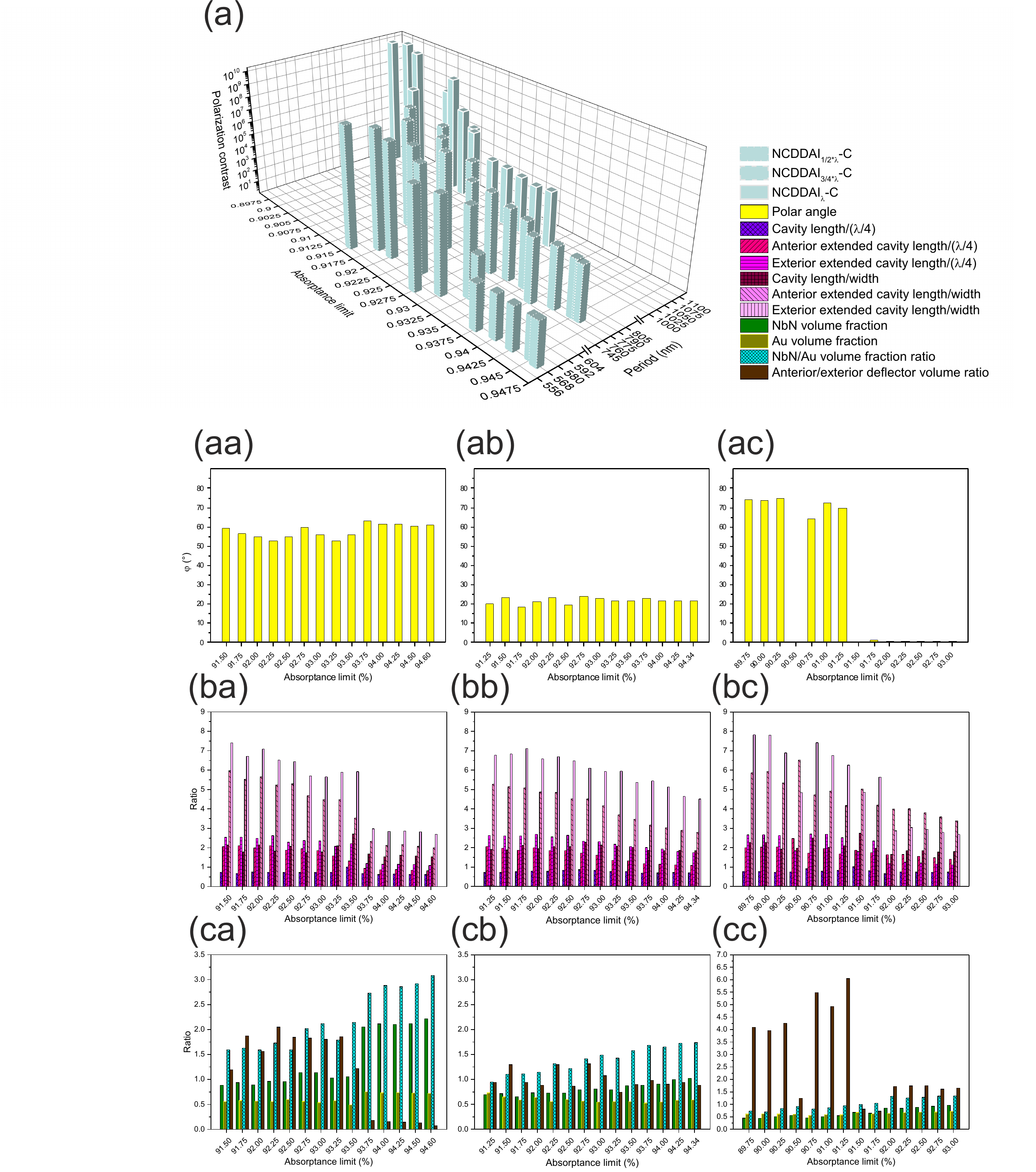}}
\caption{(a) Attained polarization contrast, optimized period and optimal polar angle as a function of conditional absorptance. Histograms indicating tendencies of (b) cavity length/$(\lambda/4)$, anterior extended cavity length/$(\lambda/4)$, exterior extended cavity length/$(\lambda/4)$, cavity length/width, anterior extended cavity length/width and exterior extended cavity length/width ratios qualifying MIM modes squeezing, (c) NbN and Au volume fraction, NbN/Au volume fraction ratio, anterior/exterior deflector volume fraction ratio, as a function of conditional absorptance, all in (a-c/a) NCDDAI$_{1/2*\lambda}$-C, (a-c/b) NCDDAI$_{3/4*\lambda}$-C, (a-c/c) NCDDAI$_{\lambda}$-C series.}
\label{figA3}
\end{figure}

In NCDDAI-C-SNSPD devices the polarization contrast rapidly and monotonously increases by decreasing the conditional absorptance, except in NCDDAI$_{\lambda}$-C, where the increase is non-monotonous (Figure~\ref{figA3}(a)). The polarization contrast increases in $[1.38*10^3-2.24*10^9]/[2.00*10^4-2.83*10^8]/[1.39*10^4-1.35*10^{10}]$ intervals in NCDDAI$_{1/2*\lambda}$-C/NCDDAI$_{3/4*\lambda}$-C/NCDDAI$_{\lambda}$-C. The period non-monotonously varies in [567.82 nm-558.56 nm]/[769.93 nm-784.41 nm]/[1042.83 nm-1058.73 nm] regions. In NCDDAI$_{1/2*\lambda}$-C the tilting non-monotonously varies in [60.89$^{\circ}$-59.57$^{\circ}$] region, while in NCDDAI$_{3/4*\lambda}$-C slightly modifies in [21.85$^{\circ}$-20.04$^{\circ}$] region, i.e. the optimal polar angle is almost independent on the conditional absorptance. The optimal polar angle exhibits a jump in [0.69$^{\circ}$-74.26$^{\circ}$] interval in NCDDAI$_{\lambda}$-C, indicating that different nanophotonical phenomena become to play below certain level of conditional absorptance (Figure~\ref{figA3}(a)/(a-c)).

The cavity lengths non-monotonously increase in $[0.62*(\lambda/4)-0.74*(\lambda/4)]/[0.71*(\lambda/4)-0.74*(\lambda/4)]/[0.74*(\lambda/4)-0.77*(\lambda/4)]$ intervals in NCDDAI$_{1/2*\lambda}$/NCDDAI$_{3/4*\lambda}$/NCDDAI$_{\lambda}$-C SNSPDs (Figure~\ref{figA3}(b)/(a-c)). The cavity parameter tendencies result in non-monotonous cavity length/width ratio, which exhibits the most well defined single maximum in NCDDAI$_{3/4*\lambda}$-C. The extended cavity length for anterior - exterior deflector side illumination exhibits almost monotonous increase from $0.81*(\lambda/4) – 1.09*(\lambda/4)$ through $2.05*(\lambda/4) - 2.54*(\lambda/4)$ in NCDDAI$_{1/2*\lambda}$-C, from $1.08*(\lambda/4) – 1.75*(\lambda/4)$ through $2.05*(\lambda/4) - 2.63*(\lambda/4)$ in NCDDAI$_{3/4*\lambda}$-C, while non-monotonous increase can be seen from $1.40*(\lambda/4) – 1.11*(\lambda/4)$ through $2.00*(\lambda/4) - 2.66*(\lambda/4)$ in NCDDAI$_{\lambda}$-C, respectively. The deflector's presence manifest itself in a well-defined almost monotonous increase in extended cavity length/width ratio in all NCDDAI-SNSPD-C. The type of MIM modes that are supported by extended cavities, when anterior side deflectors are illuminated, modifies from quarter- to half-wavelength in all NCDDAI-C. For illumination of exterior side deflectors, the MIM modes type modifies from quarter- to three-quarter-wavelength in NCDDAI$_{1/2*\lambda}$-C and NCDDAI$_{\lambda}$-C, while in NCDDAI$_{3/4*\lambda}$-C half-wavelength-scaled is the original mode. The ratio of deflectors on the anterior/exterior side of cavities increases rapidly/slowly/non-monotonously in [0.08-1.20]/[0.88-0.94] and [1.65-4.09] intervals in NCDDAI$_{1/2*\lambda}$-C, NCDDAI$_{3/4*\lambda}$-C and NCDDAI$_{\lambda}$-C, respectively (Figure~\ref{figA3}(c)/(a-c)). The NbN/Au volume fraction almost monotonously decreases in $[3.08*10^{-3}-1.59*10^{-3}]/[17.40*10^{-4}-9.47*10^{-4}]/[13.40*10^{-4}-7.30*10^{-4}]$ regions in NCDDAI$_{1/2*\lambda}$-C/NCDDAI$_{3/4*\lambda}$-C/NCDDAI$_{\lambda}$-C devices.

\subsection{NCTAI-C-SNSPD Capable of Maximizing the Polarization Contrast at Conditional Absorptance}
\begin{figure}[h]
\centerline{\includegraphics[width=0.8\columnwidth,draft=false]{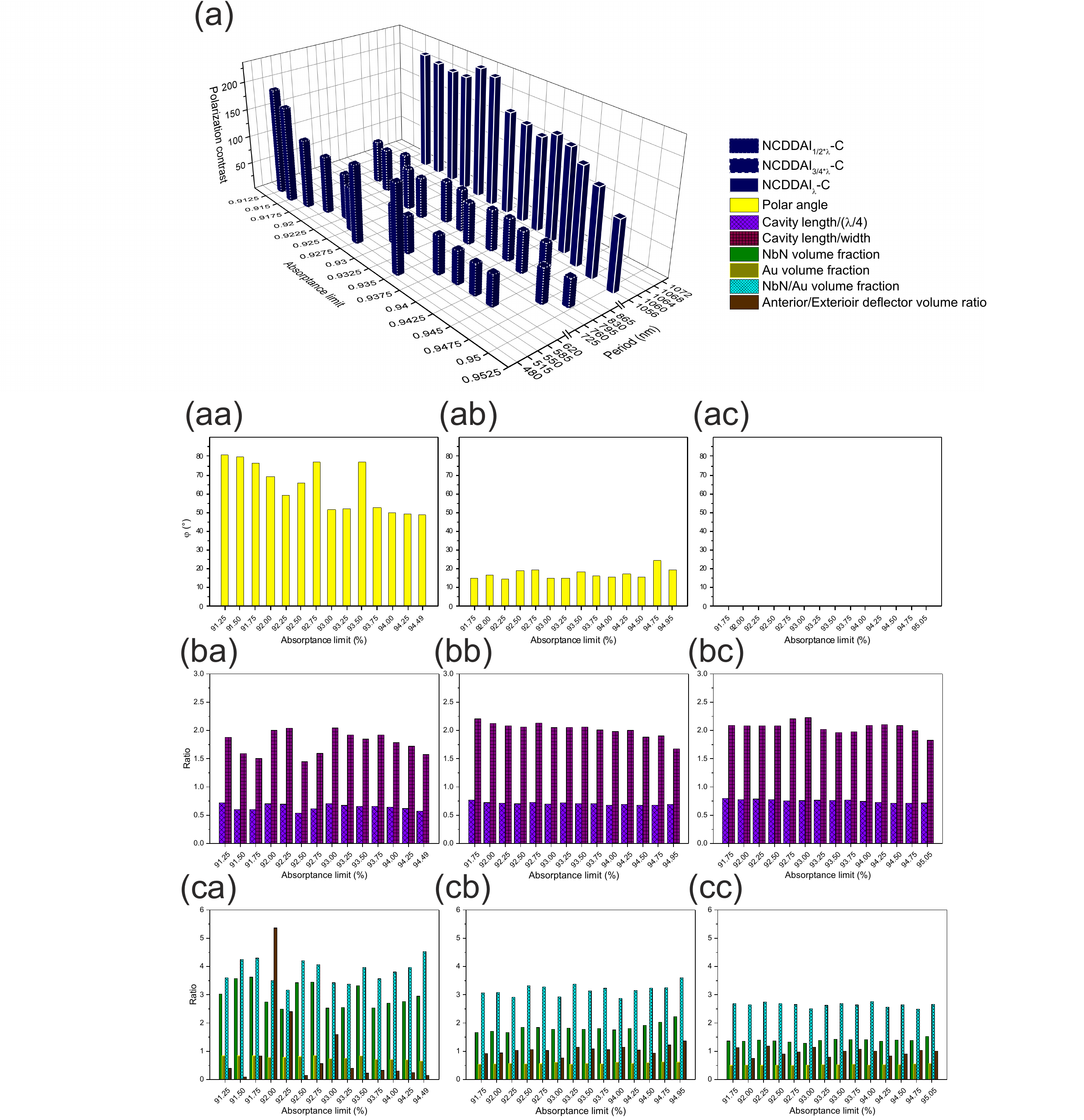}}
\caption{(a) Attained polarization contrast, optimized period and optimal polar angle as a function of conditional absorptance. Histograms indicating tendencies of (b) cavity length/$(\lambda/4)$ and cavity length/width ratios qualifying MIM modes squeezing, (c) NbN and Au volume fraction, NbN/Au volume fraction ratio, anterior/exterior deflector volume fraction ratio, all in (a-c/a) NCTAI$_{1/2*\lambda}$-C, (a-c/b) NCTAI$_{3/4*\lambda}$-C, (a-c/c) NCTAI$_{\lambda}$-C series.}
\label{figA4}
\end{figure}

In NCTAI-C-SNSPD devices the polarization contrast increases monotonously by decreasing the conditional absorptance, except in NCTAI$_{1/2*\lambda}$-C (Figure~\ref{figA4}(a)). The polarization contrast varies in $[5.53*10^1-1.91*10^2]/[6.21*10^1-7.73*10^1]/[1.55*10^2-2.16*10^2]$ intervals in NCTAI$_{1/2*\lambda}$-C/NCTAI$_{3/4*\lambda}$-C/NCTAI$_{\lambda}$-C. These results show that very low polarization contrast is achievable in presence of trenches. The period non-monotonously varies in [600.00 nm-513.75 nm]/[795.83nm-846.83 nm]/[1056.24 nm-1068.41 nm] regions. The optimal polar angle increases non-monotonously in [49.00$^{\circ}$ - 80.84$^{\circ}$] tilting interval in NCTAI$_{1/2*\lambda}$-C, indicating again that significantly different nanophotonical phenomena are at play at different conditional absorptance. In NCTAI$_{3/4*\lambda}$-C/NCTAI$_{\lambda}$-C the polar angle varies in $[19.37^{\circ}-15.09^{\circ}]/[0^{\circ}-3.41^{\circ}*10^{-3}]$, i.e. the optimal tilting is almost independent on the conditional absorptance (Figure~\ref{figA4}(a)/(a-c)). 

The cavity length exhibits non-monotonous increase in $[0.57*(\lambda/4)-0.72*(\lambda/4)]/[0.69*(\lambda/4)-0.77*(\lambda/4)]/[0.72*(\lambda/4)-0.79*(\lambda/4)]$ in NCTAI$_{1/2*\lambda}$-C/NCTAI$_{3/4*\lambda}$-C/NCTAI$_{\lambda}$-C (Figure~\ref{figA4}(b)/(c-a)). Tendencies in geometrical parameters result in non-monotonous cavity length/width ratio except in NCTAI$_{3/4*\lambda}$-C, which exhibits an almost monotonous increase. The cavity lengths are smaller than quarter-wavelength in all NCTAI-C-SNSPD devices indicating that the resonant MIM modes are strongly confined. Accordingly, quarter-wavelength type resonant MIM are supported, their wavelength gradually become similar to that of slightly squeezed SPP modes. The ratio of deflectors on the anterior/exterior side of cavities modifies non-monotonously in the interval of [0.15-0.39]/[1.36-0.92]/[1.00-1.13] in NCTAI$_{1/2*\lambda}$/NCTAI$_{3/4*\lambda}$-C/NCTAI$_{\lambda}$-C, respectively (Figure~\ref{figA4}(c)/(c-a)). The NbN/Au volume fraction ratio is non-monotonous in all NCTAI-C devices.

In summary, configurations of C-SNSPD devices, which are capable of maximizing the polarization contrast at conditional absorptance were determined for four different SNSPD device types in three periodicity intervals. Tendencies in configuration parameters ensuring polarization contrast maximization for different conditional absorptances show that in polarization contrast maximization enhancement of p-polarization specific coupling phenomena is dominant rather than squeezing of MIM modes. Based on these results the polarization contrast indicates correlation with (extended) cavity length/width, (extended) cavity length/$(\lambda/4)$ and NbN/Au volume fraction ratio parameters. The highest values achievable are on the order of $10^2-10^2-10^2/10^7-10^3-10^3/10^9-10^8-10^{10}/10^2-10^1-10^2$ in $1/2*\lambda-3/4*\lambda - \lambda$ in NCAI/NCDAI/NCDDAI/NCTAI-C-SNSPDs.
\newpage

\end{document}